%% file: main.tex
\documentclass[%
 aip,
 amsmath,amssymb,
 reprint,%
]{revtex4-1}

\usepackage{graphicx}
\usepackage{dcolumn}
\usepackage{bm}

\usepackage[utf8]{inputenc}
\usepackage[T1]{fontenc}
\usepackage{mathptmx}
\usepackage{etoolbox}

\makeatletter
\def\@email#1#2{%
 \endgroup
 \patchcmd{\titleblock@produce}
  {\frontmatter@RRAPformat}
  {\frontmatter@RRAPformat{\produce@RRAP{*#1\href{mailto:#2}{#2}}}\frontmatter@RRAPformat}
  {}{}
}%
\makeatother

\usepackage{comment}
\DeclareUnicodeCharacter{2212}{-}
\DeclareUnicodeCharacter{1E97}{t}
\DeclareUnicodeCharacter{22C5}{⋅}
\usepackage{braket}
\usepackage[hidelinks]{hyperref}
\usepackage{amsmath}
\usepackage{xcolor}
\usepackage{datetime}
\usepackage{mathtools}
\usepackage{cases}
\newcommand{\virgolette}[1]{``#1''}
\usepackage{multirow}
\usepackage{gensymb}
\usepackage{siunitx}
\usepackage{braket}
\usepackage{cases}
\newcommand\eqau{\stackrel{\mathclap{\normalfont\mbox{au}}}{=}}
\newcommand\eqlen{\stackrel{\mathclap{\normalfont\mbox{l.}}}{=}}
\newcommand\eqvel{\stackrel{\mathclap{\normalfont\mbox{v.}}}{=}}
\newcommand\bbeta{\beta_{\alpha\beta\gamma}}
\newcommand\aJ{{}^{\alpha}\!J_{\alpha\beta\gamma}}
\newcommand\bJ{{}^{\beta}\!J_{\alpha\beta\gamma}}
\newcommand\aK{{}^{\alpha}\!K_{\alpha\beta\gamma\delta}}
\newcommand\bK{{}^{\beta}\!K_{\alpha\beta\gamma\delta}}
\newcommand\oA{\hat{A}_{\alpha}}
\newcommand\oB{\hat{B}_{\beta}}
\newcommand\oC{\hat{C}_{\gamma}}
\newcommand\pll{\langle\langle}
\newcommand\prr{\rangle\rangle}
\newcommand{\omu}{\hat{\mu}}
\newcommand{\oma}{\hat{m}}
\newcommand{\oqu}{\hat{Q}}
\newcommand{\iu}{\text{i}}
\newcommand\betalendal{(-1/2) \cdot \pll \omu_{\alpha}; \omu_{\beta}, \omu_{\gamma}\prr_{\omega, \omega}}
\newcommand\betaveldal{(-1/2) \cdot [-\iu/(2\omega^{3})]\pll \omu_{\alpha}^{p}; \omu_{\beta}^{p}, \omu_{\gamma}^{p}\prr_{\omega, \omega}}
\newcommand{\lrfee}{\pll \omu_{\alpha}; \omu_{\beta} \prr_{\omega}}
\newcommand{\lrfpp}{\pll \omu_{\alpha}^{p}; \omu_{\beta}^{p} \prr_{\omega}}
\newcommand{\ga}{\alpha}
\newcommand{\gb}{\beta}
\newcommand{\gc}{\gamma}
\newcommand{\gd}{\delta}
\newcommand{\Kd}{\delta}

\usepackage{xr}
\usepackage{appendix}

\begin{document}

\preprint{AIP/123-QED}

\title[Velocity formulations for HRS-OA]{Velocity formulations for hyper-Rayleigh scattering optical activity spectroscopy: addressing the origin-dependence problem}
\author{Andrea Bonvicini}
\email{andrea.bonvicini@unamur.be}
\affiliation{Theoretical Chemistry Laboratory,
	Unit of Theoretical and Structural Physical Chemistry,
	Namur Institute of Structured Matter,
	University of Namur,
	B-5000
	Namur,
	Belgium}
\author{Sonia Coriani}
\email{soco@kemi.dtu.dk}
\affiliation{DTU Chemistry, Technical University of Denmark, Kemitorvet Bldg. 207, 2800 Kongens Lyngby, Denmark}
\author{Benoît Champagne}%
\email{benoit.champagne@unamur.be}
\affiliation{Theoretical Chemistry Laboratory,
	Unit of Theoretical and Structural Physical Chemistry,
	Namur Institute of Structured Matter,
	University of Namur,
	B-5000
	Namur,
	Belgium}

\date{\today \, $-$ \currenttime}

\begin{abstract}
The theory of hyper-Rayleigh scattering optical activity (HRS-OA) spectroscopy has previously been described within the length formulation of the pure electric-dipole and mixed (electric-dipole/magnetic-dipole and electric-dipole/electric-quadrupole) first hyperpolarizabilities required for the description of this process. In this work, we provide an alternative formulation of these pure and mixed hyperpolarizabilities. This new formulation made use of the velocity form of the electric-dipole and electric-quadrupole moment operators that enter in the quadratic response functions. A one-to-one correspondence is found for the gauge-origin shifts obtained in the two formulations. These relations ensure the origin-independence of the theory also for the velocity formulation. Furthermore, even though the basis set dependence of the velocity formulation is more significant compared to the length one, the former is origin-independent by design. This property makes it particularly suitable for calculations of HRS-OA invariants using approximated (variational or not) wavefunctions.
\end{abstract}

\maketitle

\section{Introduction} %
\label{sec:intro}
Several (chir)optical spectroscopies containing electric-dipole as well magnetic-dipole, and/or electric-quadrupole interactions are usually described within the length formulation (or length gauge) of the electric multipole (typically dipole and quadrupole) moments.\cite{Craig1998, Barron2004, Raab2004, Norman2018} Examples are
electronic circular dichroism (ECD),\cite{Power1974, Crawford2006a, Autschbach2009, Autschbach2011b, Gendron2019} 
vibrational circular dichroism (VCD),\cite{Stephens1985, Crawford2006a, Autschbach2009, Nafie2011, Coriani2011,  Shumberger2024, Shumberger2025}
optical rotation (OR),\cite{Power1971, Stephens2001, Crawford2006a, Autschbach2009, Autschbach2011b}
Rayleigh optical activity (RayOA),\cite{Andrews1980}
Raman optical activity (ROA),\cite{Barron2004, Crawford2006a, Autschbach2009, Nafie2011}
hyper-Raman optical activity (HROA),\cite{Andrews1979, Jones2024} hyper-Rayleigh scattering optical activity (HRS-OA),\cite{Bonvicini2023}
third-harmonic scattering optical activity (THS-OA),\cite{Bonvicini2023a}
two- and three-photon circular dichroism (2ECD, 3ECD),\cite{Power1975, Tinoco1975, Rizzo2006, Friese2016}
magnetochiral birifringence,\cite{Coriani2002}
circular intensity difference electric-field-induced second harmonic generation (CID-EFISHG),\cite{Lam1982, Rizzo2012}
magneto-chiral dichroism (MChD),\cite{Barron1984, Cukras2016, Atzori2021, Cukras2025}
and magnetic circular dichroism (MCD).\cite{Healy1976, Coriani1999b, Mason2007, Kjærgaard2012, Faber2020}
It is important to remember that MCD does not require a chiral molecule to be observed.\cite{Caldwell1971}
For exact (variational) wavefunctions, the length formulation of these spectroscopies can be proven to be origin-independent.\cite{Warnke2012, Rizzo2012, Bonvicini2023, Bonvicini2023a}
On the other hand, for approximated wavefunctions, and quantum chemistry calculations with finite basis sets, the length formulation provides unphysical origin-dependent results, due to the presence of magnetic-dipole (and possibly electric-quadrupole) interactions.
To overcome this issue, various strategies can be employed.
Still within the length formulation, one possible solution is to use London atomic orbitals (LAOs),\cite{London1937, Dltchfield1972, Helgaker1991} also known as gauge-invariant (or gauge-including) atomic orbitals (GIAOs),\cite{Hansen1985} which are the usual atomic orbital basis functions multiplied by a magnetic-field dependent imaginary phase factor.\cite{Jaszunski2017}
This approach has been applied to, for example,
ECD,\cite{Bak1995, Autschbach2011}
VCD,\cite{Bak1993, Cheeseman1996, Stephens2000}
OR,\cite{Cheeseman2000, Stephens2001, Ruud2002a, Stephens2005, Autschbach2011}
ROA,\cite{Helgaker1994, Liegeois2007}
2ECD,\cite{Friese2015}
CID-EFISHG,\cite{Anelli2017}
the frequency-dependent Verdet constant,\cite{Coriani2000, Krykunov2005}
and the Faraday $\mathcal{B}$ term of MCD.\cite{Coriani2000, Kjærgaard2007}
Recently, and always within the length formulation but without using LAOs, and following a method proposed by Pelloni and Lazzeretti,\cite{Pelloni2014} Caricato and collaborators presented the \virgolette{length gauge origin invariant} [LG(OI)] approach.
This method is based on the singular value decomposition and it requires the calculation of a pure electric-dipole polarizability tensor in a mixed length/velocity representation.
The LG(OI) approach has been applied for the calculation of origin-independent (individual) diagonal components of OR tensor (also known as Buckingham/Dunn optical activity tensor) on isotropic\cite{Caricato2020} and also on oriented\cite{Caricato2021} samples.
Moreover, it has also been applied to ECD spectroscopy.\cite{Niemeyer2022}
Very recently, Shumberger, Cheeseman, Caricato, and Crawford extended the LG(OI) approach to VCD.\cite{Shumberger2026}

Another alternative solution to overcome the origin-dependence issue is based on the use of the velocity formulation, which will ensure origin-independent results for both exact and approximated (variational and non-variational) wavefunctions. This has been applied to several chiroptical spectroscopies like ECD,\cite{Pecul2005a, Crawford2006a, Warnke2012}
VCD,\cite{Amos1988, Nafie1992, Ditler2022, Kumar2025}
OR,\cite{Grimme2002, Pedersen2004, Krykunov2005a, Pecul2005a, Crawford2006a, Zhang2021}
2ECD in different variants,\cite{Rizzo2006, Friese2015, Friese2016}
3ECD,\cite{Friese2016a}
and ROA.\cite{Luber2008}
In particular, for OR in the velocity representation, the presence of an unphysical behaviour in the static limit has been addressed by introducing the modified velocity gauge, which amounts to removing the unphysical static limit of the response function.\cite{Pedersen2004}
Last but not least, within the time-dependent current-density-functional theory (TDCDFT), Raimbault de Boeij, Romaniello, and Berger presented a gauge-invariant approach to ECD spectra calculation which is based on an explicit functional of the current density, instead of the particle density.\cite{Raimbault2016}

In this work, two velocity (\virgolette{velocity} and \virgolette{full-velocity}) formulations for the hyper-Rayleigh scattering optical activity (HRS-OA), which is the chiroptical version of the hyper-Rayleigh scattering (HRS) spectroscopy, are derived and illustrated for the first time. The central quantity in an HRS-OA experiment is the circular differential scattering ratio, $\Delta_{\mu}(\theta)$, a dimensionless quantity defined by\cite{Andrews1979}
\begin{equation}
\Delta_{\mu}(\theta) =
\frac{{}^{R}I_{\mu}(\theta) - {}^{L}I_{\mu}(\theta)}{{}^{R}I_{\mu}(\theta) + {}^{L}I_{\mu}(\theta)},
\label{eq:Delta_definition}	
\end{equation}
where ${}^{R/L}I_{\mu}(\theta)$ represents the intensity of the scattered light at the second harmonic obtained from right ($R$) or left ($L$) circularly polarized input photons, $\mu$ is the polarization state of the scattered photon ($\mu = \parallel, \perp$), and $\theta$ is the scattering angle which is defined by  $\cos(\theta) = -\mathbf{k} \cdot \mathbf{k}^{\prime}$, where $\mathbf{k}$ and $\mathbf{k}^{\prime}$ are the wavevectors of the incident and scattered photons, respectively. In a HRS-OA experiment, the chiral signal (i.e., the signal in the numerator of $\Delta_{\mu}$) comes from the interference between the pure electric-dipole hyperpolarizability ($\bbeta$) and four mixed hyperpolarizabilities in which one electric-dipole interaction is replaced by a magnetic-dipole interaction ($\aJ$ and $\bJ$) or an electric-quadrupole interaction ($\aK$ and $\bK$).
At the same time, the achiral signal (i.e., the signal that appears in the denominator of $\Delta_{\mu}$) is given by the interference of $\bbeta$ with itself.
By using the formalism of response theory, all these first hyperpolarizabilities are specific cases of a generic quadratic response function\cite{Norman2018}
\begin{equation}
	\langle\langle \hat{A}_{\alpha}; \hat{B}_{\beta}, \hat{C}_{\gamma} \rangle\rangle_{\omega_1, \omega_2}
	=
	\frac{1}{2}
	\sum \mathcal{P}_{-\sigma,1,2}
	\sum_{m,n}{}^{\prime}
	\frac
	{A_{\alpha}^{0m} \overline{B}_{\beta}^{mn}C_{\gamma}^{n0}}
	{(E_{m0} - E_{\sigma})(E_{n0} - E_{2})},
	\label{eq:qrf}
\end{equation}
where $\hat{A}$, $\hat{B}$ and $\hat{C}$ are generic quantum mechanical operators, $\omega_1$ and $\omega_2$ are the frequencies of the two incident photons and are associated with the operators $\hat{B}$ and $\hat{C}$, respectively, while $-\omega_{\sigma}$ is associated with $\hat{A}$, and $\omega_{\sigma} = \omega_1 + \omega_2$. In HRS-OA, $\omega_{1} = \omega_{2} = \omega$. Moreover, $E_{m0} = E_{m} - E_{0}$ and $E_{\sigma} = \hbar \omega_{\sigma}$. The notation $A_{\alpha}^{0m}\overline{B}_{\beta}^{mn}C_{\gamma}^{n0}$ is a shorthand notation for $\langle 0 |\hat{A}_{\alpha}| m\rangle \langle m | \overline{\hat{B}_{\beta}} | n\rangle \langle n |\hat{C}_{\gamma}| 0\rangle$ where $\overline{\hat{B}_{\beta}}$ is the fluctuation of the operator $\hat{B}$ and it is equal to $\hat{B}_{\beta} - \langle 0 | \hat{B}_{\beta} | 0 \rangle$. The subscripts $\alpha$, $\beta$, and $\gamma$ register a generic Cartesian axis of the microscopic frame $(x, y, z)$, $\ket{0}$ is the ground electronic state of energy $E_{0}$, while $\ket{m}$ and $\ket{n}$ are excited electronic states of energies $E_{m}$ and $E_{n}$; the prime next to the summation symbol excludes the ground state from the summation (i.e., $\ket{m}$ and $\ket{n} \neq \ket{0}$). The $1/2$ prefactor is present due to the B-convention.\cite{Stahelin1993, Reis2006}
Finally, the operator $\mathcal{P}_{-\sigma, 1, 2}$ generates the permutation of the operator-frequency pairs ($\hat{A}_{\alpha}$, $-\omega_{\sigma}$), ($\hat{B}_{\beta}$, $\omega_{1}$), and ($\hat{C}_{\gamma}$, $\omega_{2}$), thus generating a total of 6 terms.

Within the length formulation, the hyperpolarizabilities required for the molecular description of the HRS-OA spectroscopy are given by the following expressions\cite{Bonvicini2023}
\begin{align}
	\beta_{\alpha\beta\gamma}         		&\eqlen \pll \omu_{\alpha}; \omu_{\beta}, \omu_{\gamma} \prr_{\omega, \omega}, \label{eq:beta}\\
	\aJ  		&\eqlen \pll \oma_{\alpha}; \omu_{\beta}, \omu_{\gamma} \prr_{\omega, \omega}, \label{eq:aJ}\\
	\bJ   		&\eqlen
	\pll \omu_{\alpha}; \omu_{\beta}, \oma_{\gamma} \prr_{\omega, \omega}
	+
	\pll \omu_{\alpha}; \omu_{\gamma}, \oma_{\beta} \prr_{\omega, \omega}, \label{eq:bJ}\\
	\aK  &\eqlen \pll \oqu_{\alpha\delta}; \omu_{\beta}, \omu_{\gamma} \prr_{\omega, \omega}, \label{eq:aK}\\
	\bK  &\eqlen
	\pll \omu_{\alpha}; \omu_{\beta}, \oqu_{\gamma\delta} \prr_{\omega, \omega}
	+
	\pll \omu_{\alpha}; \omu_{\gamma}, \oqu_{\beta\delta} \prr_{\omega, \omega}. \label{eq:bK}
\end{align}
In particular, the left superscripts $\alpha$ and $\beta$ in $\aJ$ and $\bJ$ ($\aK$ and $\bK$) indicate whether the magnetic-dipole (electric-quadrupole) interaction is associated with the emission or absorption process, respectively.
Moreover, in eqs. \eqref{eq:beta}-\eqref{eq:bK}, the symbol $\eqlen$ is used to stress the fact that these first hyperpolarizabilities are given in the length formulation. The operators $\omu_{\alpha}$, $\oma_{\alpha}$, and $\oqu_{\alpha\beta}$ are the electric-dipole, the magnetic-dipole, and the (traceless) electric-quadrupole moment operators, respectively, whose definitions are given by the following expressions
\begin{align}
	\omu_{\alpha} 		&= \sum_{i} q_{i} \hat{r}_{i,\alpha}, \label{eq:e_dip}\\
	\oma_{\alpha} 		&= \frac{1}{2c_{0}}\sum_{i} \frac{q_{i}}{m_{i}} \epsilon_{\alpha\beta\gamma}\hat{r}_{i,\beta}\hat{p}_{i,\gamma}, \label{eq:m_dip}\\
	\oqu_{\alpha\beta} 	&=  \frac{1}{2} \sum_{i} q_{i} \left(\hat{r}_{i,\alpha}  \hat{r}_{i,\beta} -\frac{1}{3} \hat{r}_{i,\gamma}\hat{r}_{i,\gamma} \delta_{\alpha\beta} \right), \label{eq:e_quad}
\end{align}
where $q_{i}$ and $m_{i}$ are the charge and mass of the $i$-th particle (electrons and nuclei), respectively, while $\hat{r}_{i,\alpha}$ and $\hat{p}_{i,\alpha}$ are the position and linear momentum operators along the $\alpha$-direction (in the molecular Cartesian frame) for the $i$-th particle, $\delta_{\alpha\beta}$ is the Kronecker delta, $\epsilon_{\alpha\beta\gamma}$ is the Levi-Civita tensor, and $c_{0}$ is the speed of light ($\approx 137$ in atomic units).
Einstein's implicit summation over repeated indices is used in eq. \eqref{eq:m_dip} and in the following.
In this work, we made use of the position representation for the position and linear momentum operators, i.e.:\cite{Merzbacher1998}
\begin{align}
	\hat{r}_{i,\alpha} \Rightarrow \hat{r}_{i,\alpha} &\equiv r_{i,\alpha}, \label{eq:r_pos}\\
	\hat{p}_{i,\alpha} \Rightarrow \hat{p}_{i,\alpha} &\equiv -\iu \hbar \frac{\partial}{\partial r_{i,\alpha}} \label{eq:p_pos},
\end{align}
where $\iu$ is the imaginary unit. Moreover, we consider electronic wavefunctions i.e., the nuclei position and linear momenta are not quantum mechanical operators of the electronic wavefunction.
By using these definitions, for real electronic wavefunctions and away from resonances, $\bbeta$, $\aK$, and $\bK$ are real (Cartesian) tensors, while $\aJ$ and $\bJ$ are imaginary (Cartesian) tensors. Moreover, these pure and mixed first hyperpolarizabilities are symmetric in their $(\beta,\gamma)$-indices. Finally, $\aK$ is also symmetric and traceless in its $(\alpha,\delta)$-indices. These permutation and traceless symmetry properties will be preserved also in the velocity formulation derived in Section \ref{sec:velocity_formulation}.

The circular differential scattering ratio of HRS-OA, eq. \eqref{eq:Delta_definition}, can be expressed at the molecular level via
\begin{align}
	\Delta_{\parallel}(\theta) &= \frac{a + b \cos(\theta) + c \cos^{2}(\theta) + d \cos^{3}(\theta)}{f + g \cos^{2}(\theta)}, \label{eq:Delta_par}\\
	\Delta_{\perp}(\theta) &= \frac{a + (b + d) \cos(\theta) + c}{f + g}, \label{eq:Delta_per}
\end{align}
where the $a$, $b$, $c$, $d$, $f$, and $g$ terms are contributions to the scattered signal (intensity) at the second-harmonic. In particular, the $a$, $b$, $c$, and $d$ terms appear only in the numerator of eqs. \eqref{eq:Delta_par}-\eqref{eq:Delta_per} and they represent the contribution to the chiral signal while the $f$ and $g$ terms appear only in the denominator and they represent the contribution to the achiral signal. The explicit expressions of these terms are quite long and can be found in ref. \citenum{Bonvicini2023}. In this work, we will only provide a symbolic notation:
\begin{align}
	a &= \langle \operatorname{Im} \{ \beta~{}^\beta\!J  \} \rangle_{a} + |k|\langle \beta~{}^\beta\!K \rangle_{a},		\label{eq:a}\\
	b &= \langle \operatorname{Im} \{ \beta~{}^\alpha\!J \} \rangle_{b} + |k'|\langle \beta~{}^\alpha\!K \rangle_{b},	\label{eq:b}\\
	c &= \langle \operatorname{Im} \{ \beta~{}^\beta\!J  \} \rangle_{c} + |k|\langle \beta~{}^\beta\!K \rangle_{c},		\label{eq:c}\\
	d &= |k'| \langle \beta~{}^\alpha\!K \rangle_{d},										\label{eq:d}\\
	f &= \langle \beta \beta \rangle_{f},													\label{eq:f}\\
	g &= \langle \beta \beta \rangle_{g},													\label{eq:g}
\end{align}
where $\operatorname{Im}\{z\}$ stands for the imaginary part of a generic complex number $z$ and, for example, the notation $\langle \operatorname{Im}(\beta~{}^\beta\!J) \rangle_{a}$ indicates the sum of all the molecular invariants (that derives from the rotational averaging procedure) that are given by the interference between $\bbeta$ with the imaginary part of $\bJ$ and the suffix $a$ is used to specify that these molecular invariants belong to $a$, i.e., in general, $\langle \operatorname{Im}(\beta {}^{\beta}J) \rangle_{a} \neq \langle \operatorname{Im}(\beta {}^{\beta}J) \rangle_{c}$. As it will be discussed later, this formulation is origin-independent for exact wavefunctions or for approximated variational wavefunctions in the limit of a complete basis set. However, for approximated wavefunctions (e.g., variational wavefunctions with finite basis set) this formulation is origin-dependent. The aim of this work is to provide an origin-independent formulation of HRS-OA spectroscopy which is based on the use of the velocity formulation of the five first hyperpolarizabilities listed in eqs. \eqref{eq:beta}-\eqref{eq:bK}.

\section{From the length to the velocity formulation} %
\label{sec:velocity_formulation}

\subsection{General considerations} %
The velocity formulation of the five first hyperpolarizabilities that appear in HRS-OA, eqs. \eqref{eq:beta}-\eqref{eq:bK}, is based on the use of the following hypervirial relations for linear and quadratic response functions\cite{Olsen1995a, Coriani2002, Helgaker2012, Pedersen2017a}
\begin{widetext}
\begin{align}
\langle 0 | \hat{A}_{\alpha} | n \rangle &= \frac{1}{\omega_{n0}} \langle 0 | [\hat{A}_{\alpha},\hat{H}_{\text{mol}}] | n \rangle,
\label{eq:hyp_0}\\
\pll \hat{A}_{\alpha};\hat{B}_{\beta}\prr_{\omega_{1}} &= \frac{1}{\omega_{1}}
\left(
\pll [\hat{A}_{\alpha},\hat{H}_{\text{mol}}];\hat{B}_{\beta} \prr_{\omega_{1}}
+
\langle 0 | [\hat{A}_{\alpha},\hat{B}_{\beta}] | 0 \rangle
\right),
\label{eq:hyp_1}
\\
\pll \hat{A}_{\alpha};\hat{B}_{\beta},\hat{C}_{\gamma} \prr_{\omega_{1},\omega_{2}} &= \frac{1}{(\omega_{1} + \omega_{2})}
\left(
\pll [\hat{A}_{\alpha},\hat{H}_{\text{mol}}];\hat{B}_{\beta},\hat{C}_{\gamma} \prr_{\omega_{1},\omega_{2}}
+
\pll [\hat{A}_{\alpha},\hat{B}_{\beta}]; \hat{C}_{\gamma}\prr_{\omega_{2}}
+
\pll [\hat{A}_{\alpha},\hat{C}_{\gamma}]; \hat{B}_{\beta}\prr_{\omega_{1}}
\right),
\label{eq:hyp_2}
\end{align}
\end{widetext}
where $\hat{H}_{\text{mol}}$ is the molecular electronic Hamiltonian for which we assume that we know the exact ground and excited state electronic wavefunctions. In particular, because the first hyperpolarizabilities are quadratic response functions, the derivation of the velocity formulation is based on the use of eq. \eqref{eq:hyp_2}. Moreover, for a complete derivation, the following commutators, derived assuming electronic wavefunctions, are required (in atomic units, $\hbar=m_{e}=e=1$):\cite{Rizzo2003}
\begin{align}
[\omu_{\alpha}, \hat{H}_{\text{mol}}] 		&= \iu \omu_{\alpha}^{p}, \label{eq:comm_mu_H}\\
[\omu_{\alpha}, \omu_{\beta}] 				&= [\omu_{\alpha}, \oqu_{\beta\gamma}] = 0, \label{eq:comm_mu_mu}\\
[\omu_{\alpha}, \omu_{\beta}^{p}] 			&= \iu N_{e} \delta_{\alpha\beta}, \label{eq:comm_mu_mup}\\
[\omu_{\alpha}, \oma_{\beta}] 				&=  -\frac{\iu}{2c_{0}}\epsilon_{\alpha\beta c} \omu_{c}, \label{eq:comm_mu_m}\\
[\oqu_{\alpha\beta}, \hat{H}_{\text{mol}}] &= \iu \oqu_{\alpha\beta}^{p}, \label{eq:comm_Q_H}\\
[\omu_{\alpha}, \iu\oqu_{\beta\gamma}^{p}] &=
\frac{1}{2}
\left(
\omu_{\beta} \delta_{\alpha\gamma}
+
\omu_{\gamma} \delta_{\alpha\beta}
\right)
-
\frac{1}{3}
\omu_{\alpha}
\delta_{\beta\gamma}
\label{eq:comm_mu_iQp}\\
[\hat{Q}_{\alpha\beta}, \iu \omu_{\gamma}^{p}] &=
\frac{1}{2}
\left(
\omu_{\alpha} \delta_{\beta\gamma}
+
\omu_{\beta} \delta_{\alpha\gamma}
\right)
-\frac{1}{3}
\omu_{\gamma} \delta_{\alpha\beta}
\label{eq:comm_Q_imup}
\end{align}
where $N_{e}$ is the total number of electrons in the (molecular) system, $\omu_{\alpha}^{p}$ and $\oqu_{\alpha\beta}^{p}$ are the electric-dipole and traceless electric-quadrupole moment operators, respectively, written in the velocity form
\begin{align}
\omu_{\alpha}^{p}
&= \sum_{i} \frac{q_{i}}{m_{i}} \hat{p}_{i,\alpha},
\label{eq:e_dip_p}\\
\oqu_{\alpha\beta}^{p}
&=
\frac{1}{2}
\sum_{i}
\frac{q_{i}}{m_{i}}
\left[
\hat{r}_{i,\alpha} \hat{p}_{i,\beta}
+
\hat{p}_{i,\alpha} \hat{r}_{i,\beta}
-\frac{1}{3}
\left( \hat{r}_{i,\gamma} \hat{p}_{i,\gamma} + \hat{p}_{i,\gamma} \hat{r}_{i,\gamma} \right) \delta_{\alpha\beta}
\right] \label{eq:e_quad_p}.
\end{align}
Proofs of the commutators in eqs. \eqref{eq:comm_mu_H}-\eqref{eq:comm_Q_imup} are given in the Supporting Information file (Section \ref{sec:commutators}).

\subsection{The velocity formulation} %
By using eqs. \eqref{eq:hyp_1}-\eqref{eq:hyp_2}, and the commutators in eqs. \eqref{eq:comm_mu_H}-\eqref{eq:comm_Q_imup}, one can derive the velocity forms for the five first hyperpolarizabilities listed in eqs. \eqref{eq:beta}-\eqref{eq:bK}. These are given by
\begin{widetext}
\begin{align}
	\begin{split}
		\bbeta \eqvel
		&
		-\frac{\iu}{2\omega^{3}} \pll \omu_{\alpha}^{p}; \omu_{\beta}^{p}, \omu_{\gamma}^{p} \prr_{\omega, \omega}
	\end{split}
	\label{eq:beta_velocity}\\
	\begin{split}
		\aJ \eqvel
		&
		-\frac{1}{\omega^{2}}
		\pll \oma_{\alpha}; \omu_{\beta}^{p}, \omu_{\gamma}^{p}  \prr_{\omega,\omega}
		+\frac{\iu}{2c_{0}\omega}
		(
		\epsilon_{\gamma\alpha c} \pll \omu_{\beta}; \omu_{c} \prr_{\omega}
		+
		\epsilon_{\beta\alpha c} \pll \omu_{\gamma}; \omu_{c} \prr_{\omega}
		)
	\end{split}
	\label{eq:aJ_velocity}\\
	\begin{split}
		\bJ \eqvel
		&
		+\frac{1}{2\omega^{2}}
		\left(\vphantom{\frac{1}{1}}
		\pll \omu_{\alpha}^{p}; \omu_{\beta}^{p}, \oma_{\gamma} \prr_{\omega,\omega}
		+
		\pll \omu_{\alpha}^{p}; \omu_{\gamma}^{p}, \oma_{\beta} \prr_{\omega,\omega}
		\right)
		-\frac{\iu}{4c_{0}\omega}
		\left(\vphantom{\frac{1}{1}}
		\epsilon_{\alpha\gamma c} \pll \omu_{c}; \omu_{\beta} \prr_{\omega}
		+
		\epsilon_{\alpha\beta c} \pll \omu_{c}; \omu_{\gamma} \prr_{\omega}
		\right)
	\end{split}
	\label{eq:bJ_velocity}\\
	\begin{split}
		\aK \eqvel
		&
		-\frac{\iu}{2\omega^{3}}
		\pll
		\oqu_{\alpha\delta}^{p}; \omu_{\beta}^{p}, \omu_{\gamma}^{p} 
		\prr_{\omega, \omega}
		\\ &
		-\frac{1}{12\omega^{2}}
		\biggl[
		3
		\biggl(
		\pll
		\omu_{\alpha}; \omu_{\beta}
		\prr_{\omega}
		\delta_{\gamma\delta}
		+
		\pll
		\omu_{\delta}; \omu_{\beta}
		\prr_{\omega}
		\delta_{\gamma\alpha}
		+
		\pll
		\omu_{\alpha}; \omu_{\gamma}
		\prr_{\omega}
		\delta_{\beta\delta}
		+
		\pll
		\omu_{\delta}; \omu_{\gamma}
		\prr_{\omega}
		\delta_{\beta\alpha}
		\biggr)
		-4
		\pll
		\omu_{\gamma}; \omu_{\beta}
		\prr_{\omega}
		\delta_{\alpha\delta}
		\biggr]
	\end{split}
	\label{eq:aK_velocity}\\
	\begin{split}
	\bK \eqvel
	&
	-\frac{\iu}{2\omega^{3}}
	\left(
	\pll \omu_{\alpha}^{p}; \omu_{\beta}^{p}, \oqu_{\gamma\delta}^{p} \prr_{\omega,\omega}
	+
	\pll \omu_{\alpha}^{p}; \omu_{\gamma}^{p}, \oqu_{\beta\delta}^{p} \prr_{\omega,\omega}
	\right)\\&
	+ \frac{1}{12\omega^{2}}
	\biggr[
	-3
	\left( \vphantom{\frac{1}{1}}
	\pll \omu_{\beta}; \omu_{\delta} \prr_{\omega} \delta_{\alpha\gamma}
	+
	\pll \omu_{\gamma}; \omu_{\delta} \prr_{\omega} \delta_{\alpha\beta}
	+
	2
	\pll \omu_{\beta}; \omu_{\gamma} \prr_{\omega} \delta_{\alpha\delta}
	\right)
	+ 2
	\left(\vphantom{\frac{1}{1}}
	\pll \omu_{\beta}; \omu_{\alpha} \prr_{\omega} \delta_{\gamma\delta}
	+
	\pll \omu_{\gamma}; \omu_{\alpha} \prr_{\omega} \delta_{\beta\delta}
	\right)
	\\&
	+ 12
	\pll \omu_{\alpha}; \omu_{\delta}\prr_{2\omega} \delta_{\beta\gamma}
	+ 2
	\vphantom{\frac{1}{1}}
	\pll \omu_{\alpha}; \omu_{\gamma} \prr_{2\omega} \delta_{\beta\delta}
	+ 2
	\pll \omu_{\alpha}; \omu_{\beta} \prr_{2\omega} \delta_{\gamma\delta}
	\biggr]
\end{split}
	\label{eq:bK_velocity}
\end{align}
\end{widetext}
The step-by-step derivation of the above expressions is given in the Supporting Information file (Section \ref{sec:len_vel_derivation}).
It is important to notice that eq. \eqref{eq:beta_velocity} has been already obtained in a previous work concerning the comparison between the length and the velocity formulations of the pure electric-dipole polarizability $\alpha_{\alpha\beta}(-\omega;\omega)$, and the second-harmonic generation first hyperpolarizability $\bbeta(-2\omega;\omega,\omega)$.\cite{Bonvicini2025a}
In particular, one can notice that the first term in the rhs of eqs. \eqref{eq:beta_velocity}, \eqref{eq:aK_velocity}, and \eqref{eq:bK_velocity} contain the same prefactor, $-\iu/(2\omega^{3})$. This prefactor is rather important for the following discussion about the origin-dependence. One can notice that in the velocity formulation, the computation of $\aJ$, $\bJ$, $\aK$, and $\bK$ is computationally more expensive because it requires the linear response function $\pll \omu_{\alpha}; \omu_{\beta} \prr_{\omega}$ (which is minus the pure electric-dipole linear polarizability $\alpha_{\alpha\beta}$). Moreover, for $\bK$, it is also necessary to compute the pure electric-dipole linear response function at the second-harmonic frequency, i.e., $\pll \omu_{\alpha}; \omu_{\beta} \prr_{2\omega}$.

\vspace{2.0cm}
\subsubsection{The full velocity formulation}
Because of the presence of the pure electric-dipole linear response function in the length form, $\pll \omu_{\alpha}; \omu_{\beta} \prr_{\omega}$, in eqs. \eqref{eq:aJ_velocity}-\eqref{eq:bK_velocity}, one can consider the latters as not being in a \textit{full} velocity form, but in a mixed velocity-length form. A full velocity formulation can be obtained by
using the following relation for the pure electric-dipole linear response function\cite{Bonvicini2025a}
\begin{align}
	\pll \omu_{\alpha}; \omu_{\beta} \prr_{\omega}
	=
	\frac{1}{\omega^{2}}
	\left(
	\pll \omu_{\alpha}^{p}; \omu_{\beta}^{p} \prr_{\omega}
	+
	N_{e}
	\delta_{\alpha\beta}
	\right),
	\label{eq:alpha_velocity}
\end{align}
whose derivation is given in the Supporting Information file (Section \ref{sec:len_vel_derivation}). For $\bK$, it is also necessary to consider the velocity formulation of $\pll \omu_{\alpha}; \omu_{\beta} \prr_{2\omega}$. By using the rhs eq. \eqref{eq:alpha_velocity} in eqs. \eqref{eq:aJ_velocity}-\eqref{eq:bK_velocity}, one can obtain the full velocity formulation of the mixed first hyperpolarizabilities $\aJ$, $\bJ$, $\aK$, and $\bK$:
\begin{widetext}
\begin{align}
	\begin{split}
		\aJ \eqvel
		&
		-\frac{1}{\omega^{2}}
		\pll \oma_{\alpha}; \omu_{\beta}^{p}, \omu_{\gamma}^{p}  \prr_{\omega,\omega}
		+\frac{\iu}{2c_{0}\omega^{3}}
		\left(
		\vphantom{\frac{1}{1}}
		\epsilon_{\gamma\alpha c} \pll \omu_{\beta}^{p}; \omu_{c}^{p} \prr_{\omega}
		+
		\epsilon_{\beta\alpha c} \pll \omu_{\gamma}^{p}; \omu_{c}^{p} \prr_{\omega}
		\right)
	\end{split}
	\label{eq:aJ_velocity_full}\\
	\begin{split}
		\bJ \eqvel
		&
		+\frac{1}{2\omega^{2}}
		\left(\vphantom{\frac{1}{1}}
		\pll \omu_{\alpha}^{p}; \omu_{\beta}^{p}, \oma_{\gamma} \prr_{\omega,\omega}
		+
		\pll \omu_{\alpha}^{p}; \omu_{\gamma}^{p}, \oma_{\beta} \prr_{\omega,\omega}
		\right)
		-\frac{\iu}{4c_{0}\omega^{3}}
		\left(\vphantom{\frac{1}{1}}
		\epsilon_{\alpha\gamma c} \pll \omu_{c}^{p}; \omu_{\beta}^{p} \prr_{\omega}
		+
		\epsilon_{\alpha\beta c} \pll \omu_{c}^{p}; \omu_{\gamma}^{p} \prr_{\omega}
		\right)
	\end{split}
	\label{eq:bJ_velocity_full}\\
	\begin{split}
		\aK \eqvel
		&
		-\frac{\iu}{2\omega^{3}}
		\pll
		\oqu_{\alpha\delta}^{p}; \omu_{\beta}^{p}, \omu_{\gamma}^{p} 
		\prr_{\omega, \omega}
		\\ &
		-\frac{1}{12\omega^{4}}
		\biggl[
		3
		\left(
		\pll
		\omu_{\alpha}^{p} ; \omu_{\beta}^{p}
		\prr_{\omega}
		\delta_{\gamma\delta}
		+
		\pll
		\omu_{\delta}^{p} ; \omu_{\beta}^{p}
		\prr_{\omega}
		\delta_{\gamma\alpha}
		+
		\pll
		\omu_{\alpha}^{p} ; \omu_{\gamma}^{p}
		\prr_{\omega}
		\delta_{\beta\delta}
		+
		\pll
		\omu_{\delta}^{p} ; \omu_{\gamma}^{p}
		\prr_{\omega}
		\delta_{\beta\alpha}
		\right)
		-4
		\pll
		\omu_{\gamma}^{p} ; \omu_{\beta}^{p}
		\prr_{\omega}
		\delta_{\alpha\delta}
		\\ &
		+ N_{e}
		\biggl(
		6 \delta_{\alpha\beta}\delta_{\gamma\delta}
		+6 \delta_{\alpha\gamma}\delta_{\beta\delta}
		-4 \delta_{\alpha\delta}\delta_{\beta\gamma}
		\biggr)
		\biggr]
	\end{split}
	\label{eq:aK_velocity_full}\\
\begin{split}
	\bK \eqvel
	&
	-\frac{\iu}{2\omega^{3}}
	\left(
	\pll \omu_{\alpha}^{p}; \omu_{\beta}^{p}, \oqu_{\gamma\delta}^{p} \prr_{\omega,\omega}
	+
	\pll \omu_{\alpha}^{p}; \omu_{\gamma}^{p}, \oqu_{\beta\delta}^{p} \prr_{\omega,\omega}
	\right)
	\\ &
	+\frac{1}{24\omega^{4}}
	\biggl[
-  6   \pll \omu_{\beta}^{p} ; \omu_{\delta}^{p} \prr_{\omega}  \delta_{\alpha\gamma} 
-  6   \pll \omu_{\gamma}^{p} ; \omu_{\delta}^{p} \prr_{\omega}  \delta_{\alpha\beta}
-  12  \pll \omu_{\beta}^{p} ; \omu_{\gamma}^{p} \prr_{\omega}  \delta_{\alpha\delta} 
+  4   \pll \omu_{\alpha}^{p} ; \omu_{\beta}^{p} \prr_{\omega}  \delta_{\gamma\delta} 
+  4   \pll \omu_{\alpha}^{p} ; \omu_{\gamma}^{p} \prr_{\omega}  \delta_{\beta\delta} 
\\ &
+  6   \pll \omu_{\alpha}^{p} ; \omu_{\delta}^{p} \prr_{2\omega}  \delta_{\beta\gamma}
+      \pll \omu_{\alpha}^{p} ; \omu_{\gamma}^{p} \prr_{2\omega}  \delta_{\beta\delta} 
+      \pll \omu_{\alpha}^{p} ; \omu_{\beta}^{p} \prr_{2\omega}  \delta_{\gamma\delta} 
- N_{\text{e}}
\biggl(
        \delta_{\alpha\beta}  \delta_{\gamma\delta} 
+       \delta_{\alpha\gamma}  \delta_{\beta\delta} 
+  6    \delta_{\alpha\delta}  \delta_{\beta\gamma} 
\biggr)
\biggr]
\end{split}
\label{eq:bK_velocity_full}
\end{align}
\end{widetext}
In deriving eqs. \eqref{eq:aJ_velocity_full}-\eqref{eq:bK_velocity_full}, we made use of the the anti-symmetry of the Levi-Civita tensor (i.e., $\epsilon_{ijk} = - \epsilon_{ikj}$) together with the symmetry of the Kronecker delta tensor (i.e., $\delta_{ij} = \delta_{ji}$). Note that the velocity formulation for $\bbeta$, eq. \eqref{eq:beta_velocity}, is already in its full velocity form. For the mixed first hyperpolarizabilities, one can use eqs. \eqref{eq:aJ_velocity}-\eqref{eq:bK_velocity} or eqs. \eqref{eq:aJ_velocity_full}-\eqref{eq:bJ_velocity_full}. The two velocity formulations are irrelevant for what concerns the origin-independence of the tensors, as explained in the next Section \ref{sec:origin_dependence}. For the full velocity formulation of $\aK$ and $\bK$, eqs. \eqref{eq:aK_velocity_full} and \eqref{eq:bK_velocity_full}, one can notice the presence of terms containing the number of electron factor, $N_{\text{e}}$.
From a computational point of view, the difference between the length, the velocity, and the full velocity formulation is summarized in Table \ref{tab:computational_details}.
\begin{table*}[ht!]
	\caption{\label{tab:computational_details} List of the quantities required to calculate the five pure and mixed first hyperpolarizabilities in the length, velocity, and full-velocity formulations of HRS-OA.}
	\renewcommand*{\arraystretch}{1.5}
	\begin{ruledtabular}
		\begin{tabular}{llll}
			{}       & Length                                                                       & Velocity                                                                                                                                                                                      & Full-velocity                                                                                                                                                                                                                 \\ \hline
			$\bbeta$ & $\pll \omu_{\alpha}; \omu_{\beta}, \omu_{\gamma} \prr_{\omega,\omega}$       & $\pll \omu_{\alpha}^{p}; \omu_{\beta}^{p}, \omu_{\gamma}^{p} \prr_{\omega,\omega}$                                                                                                            & $\pll \omu_{\alpha}^{p}; \omu_{\beta}^{p}, \omu_{\gamma}^{p} \prr_{\omega,\omega}$                                                                                                                                            \\
			$\aJ$    & $\pll \oma_{\alpha}; \omu_{\beta}, \omu_{\gamma} \prr_{\omega,\omega}$       & $\pll \oma_{\alpha}; \omu_{\beta}^{p}, \omu_{\gamma}^{p} \prr_{\omega,\omega}$, $\pll \omu_{\alpha}; \omu_{\beta} \prr_{\omega}$                                                              & $\pll \oma_{\alpha}; \omu_{\beta}^{p}, \omu_{\gamma}^{p} \prr_{\omega,\omega}$, $\pll \omu_{\alpha}^{p}; \omu_{\beta}^{p} \prr_{\omega}$                                                                                      \\
			$\bJ$    & $\pll \omu_{\alpha}; \omu_{\beta}, \oma_{\gamma} \prr_{\omega,\omega}$       & $\pll \omu_{\alpha}^{p}; \omu_{\beta}^{p}, \oma_{\gamma} \prr_{\omega,\omega}$, $\pll \omu_{\alpha}; \omu_{\beta} \prr_{\omega}$                                                              & $\pll \omu_{\alpha}^{p}; \omu_{\beta}^{p}, \oma_{\gamma} \prr_{\omega,\omega}$, $\pll \omu_{\alpha}^{p}; \omu_{\beta}^{p} \prr_{\omega}$                                                                                      \\
			$\aK$    & $\pll \oqu_{\alpha\delta}; \omu_{\beta}, \omu_{\gamma} \prr_{\omega,\omega}$ & $\pll \oqu_{\alpha\delta}^{p}; \omu_{\beta}^{p}, \omu_{\gamma}^{p} \prr_{\omega,\omega}$, $\pll \omu_{\alpha}; \omu_{\beta} \prr_{\omega}$                                                    & $\pll \oqu_{\alpha\delta}^{p}; \omu_{\beta}^{p}, \omu_{\gamma}^{p} \prr_{\omega,\omega}$, $\pll \omu_{\alpha}^{p}; \omu_{\beta}^{p} \prr_{\omega}$, $N_{\text{e}}$                                                            \\
			$\bK$    & $\pll \omu_{\alpha}; \omu_{\beta}, \oqu_{\gamma\delta} \prr_{\omega,\omega}$ & $\pll \omu_{\alpha}^{p}; \omu_{\beta}^{p}, \oqu_{\gamma\delta}^{p} \prr_{\omega,\omega}$, $\pll \omu_{\alpha}; \omu_{\beta} \prr_{\omega}$, $\pll \omu_{\alpha}; \omu_{\beta} \prr_{2\omega}$ & $\pll \omu_{\alpha}^{p}; \omu_{\beta}^{p}, \oqu_{\gamma\delta}^{p} \prr_{\omega,\omega}$, $\pll \omu_{\alpha}^{p}; \omu_{\beta}^{p} \prr_{\omega}$, $\pll \omu_{\alpha}^{p}; \omu_{\beta}^{p} \prr_{2\omega}$, $N_{\text{e}}$
		\end{tabular}
	\end{ruledtabular}
\end{table*}
%

\section{The origin-dependence} %
\label{sec:origin_dependence}

Given a gauge-origin $\mathbf{O}$ of the molecular Cartesian frame, one can translate it in space according to the following linear transformation:
\begin{equation}
	\mathbf{O} \rightarrow \mathbf{O}^{\prime} = \mathbf{O} + \mathbf{R}
	\label{eq:O_new}
\end{equation}
where $\mathbf{R}$ is an arbitrary Cartesian vector. From here on, the arrow pointing from the left to the right, \virgolette{$\rightarrow$}, and the \virgolette{$\prime$} symbol indicate the change in the gauge-origin (from $\mathbf{O}$ to $\mathbf{O}^{\prime}$).

In the next subsections, the problem of origin dependence is described at different levels. We start from the effect on the quantum mechanical operators, then we discuss these changes on the different types of molecular integrals, and then on the molecular response tensors. At the end, we show how the experimental observables (the scattered intensities in HRS-OA experiments) are unaffected by these origin-dependences, i.e., for exact wavefunctions, the theory is valid, i.e., the expressions for the scattered intensities are origin-independent.

\subsection{In the quantum mechanical operators} %
In the position representation of quantum mechanics [eqs. \eqref{eq:r_pos}-\eqref{eq:p_pos}], the change in the gauge-origin $\mathbf{O}$, eq. \eqref{eq:O_new}, affects the position and the linear momentum operators for of the $i$-th particle in the following way
\begin{align}
	\hat{r}_{i,\alpha} 	&\rightarrow \hat{r}_{i,\alpha}^{\prime} = \hat{r}_{i,\alpha} - R_{\alpha},\\
	\hat{p}_{i,\alpha} 	&\rightarrow \hat{p}_{i,\alpha}^{\prime} = \hat{p}_{i,\alpha}.
\end{align}
This means that, in the position representation, any change in the gauge-origin $\mathbf{O}$ affects the position operator but not the linear momentum operator. This has important consequences for the electric-dipole, magnetic-dipole and electric-quadrupole operators\cite{Barron2004,Bonvicini2023}
\begin{widetext}
\begin{align}
\omu_{\alpha}
&\rightarrow \omu_{\alpha}^{\prime}
= \omu_{\alpha} - R_{\alpha}q_{\text{mol}},
\label{eq:mu_gauge}
\\
\omu_{\alpha}^{p}
&\rightarrow \omu_{\alpha}^{p\prime}
= \vphantom{\frac{1}{1}}\omu_{\alpha}^{p},
\label{eq:mu_p_gauge}
\\
\oma_{\alpha}
&\rightarrow \oma_{\alpha}^{\prime}
= \oma_{\alpha} - \frac{1}{2c_{0}}\epsilon_{\alpha b c} R_{b} \omu_{c}^{p},
\label{eq:m_gauge}
\\
\begin{split}
\oqu_{\alpha\beta}
&\rightarrow \oqu_{\alpha\beta}^{\prime}
= \oqu_{\alpha\beta}
-\frac{1}{2} R_{\alpha} \omu_{\beta}
-\frac{1}{2} R_{\beta}  \omu_{\alpha}
+\frac{1}{3} R_{c}      \omu_{c}		\delta_{\alpha\beta}
+\frac{1}{2} R_{\alpha} R_{\beta} q_{\text{mol}}
-\frac{1}{6} R_{\gamma}R_{\gamma}\delta_{\alpha\beta}q_{\text{mol}}
\label{eq:Q_gauge}
\end{split}\\
\oqu_{\alpha\beta}^{p}
&\rightarrow \oqu_{\alpha\beta}^{p\prime} = \oqu_{\alpha\beta}^{p}
-\frac{1}{2} R_{\alpha}\omu_{\beta}^{p}
-\frac{1}{2} R_{\beta}\omu_{\alpha}^{p}
+\frac{1}{3} R_{c}\omu_{c}^{p} \delta_{\alpha\beta}, \label{eq:Q_p_gauge}
\end{align}
\end{widetext}
where $q_{\text{mol}}$ is the total charge of the molecule,
\begin{equation}
q_{\text{mol}} = \sum_{i} q_{i}.
\end{equation}
From eq. \eqref{eq:mu_p_gauge}, one can notice that $\omu_{\alpha}^{p}$, as $\hat{p}_{i,\alpha}$, is also origin-independent. Moreover, except for the last two terms in the rhs of eq. \eqref{eq:Q_gauge}, the gauge-origin dependences of $\oqu_{\alpha\beta}$ and $\oqu_{\alpha\beta}^{p}$ are structurally very similar. However, as explained in the following subsection, the last two terms in eq. \eqref{eq:Q_gauge}, are \textit{irrelevant} for transition properties like the first hyperpolarizabilities $\aK$ and $\bK$.

\subsection{In the molecular integrals} %
Given a generic quantum mechanical operator, $\hat{\Omega}$, its shift upon the change in the gauge-origin reads:
\begin{equation}
\hat{\Omega} \rightarrow \hat{\Omega}^{\prime} = \hat{\Omega} + \hat{X} + B = \hat{A} + B
\label{eq:generic_gauge_origin}
\end{equation}
where $\hat{A} (\equiv \hat{\Omega} + \hat{X})$ is a quantum mechanical operator, while $B$ is not. Depending on the operator, $\hat{X}$ can be zero, and the same is also true for $B$. At this point, for permanent, $\braket{n|\hat{\Omega}|n}$, transition, $\braket{m|\hat{\Omega}|n}$ with $n \neq m$, and fluctuation, $\braket{m|\overline{\hat{\Omega}}|n} \equiv \braket{m|\hat{\Omega}|n} - \delta_{mn}\braket{0|\hat{\Omega}|0}$, integrals, the effects of eq. \eqref{eq:generic_gauge_origin} on them are:
%
\begin{align}
\braket{n|\hat{\Omega}|n}
\rightarrow
\braket{n|\hat{\Omega}^{\prime}|n}
&=
\braket{n|\hat{A} + B |n}
\nonumber
\\
&=
\braket{n|\hat{A}|n}
+
\braket{n|B|n}
\nonumber
\\ &=
\braket{n|\hat{A}|n}
+
B
\label{eq:permanent_int_gauge}\\
\braket{n|\hat{\Omega}|m}
\rightarrow
\braket{n|\hat{\Omega}^{\prime}|m}
&=
\braket{n|\hat{A} + B |m}
\nonumber
\\
&=
\braket{n|\hat{A}|m}
+
\braket{n|B|m}
\nonumber
\\
&=
\braket{n|\hat{A}|m}, \qquad \text{with } n \neq m
\label{eq:transition_int_gauge}\\
\braket{n|\overline{\hat{\Omega}}|m}
\rightarrow
\braket{n|\overline{\hat{\Omega}^{\prime}}|m}
&=
\braket{n|\overline{\hat{A} + B}|m}
\nonumber
\\
&=
\braket{n|\hat{A} + B|m}
-
\delta_{nm}
\braket{0|\hat{A} + B|0}
\nonumber
\\
&=
\braket{n|\hat{A}|m}
-
\delta_{nm}
\braket{0|\hat{A}|0}
\label{eq:fluctuation_int_gauge}
\end{align}
%
From eq. \eqref{eq:permanent_int_gauge}, one can notice the presence of $B$ at the new gauge for permanent moment integrals. This explain why, only the first nonvanishing permanent moment is origin-invariant.\cite{Barron2004, Raab2004} On the other hand, from the rhs of eqs. \eqref{eq:transition_int_gauge}-\eqref{eq:fluctuation_int_gauge}, the presence of $B$ in transition and fluctuation integrals is null. These considerations are important for the following discussion about the first hyperpolarizabilities and the change in the gauge-origin.

\subsection{In the molecular response tensors} %
In general, a molecular response tensor (like any response function) is affected by a change in the gauge-origin as long as the operators contained in it are affected by this change.
However, as already explained in ref. \citenum{Bonvicini2023}, the pure electric-dipole first hyperpolarizability, $\bbeta$, written in the length formulation, eq \eqref{eq:beta}, is origin-independent for any system: neutral, dipolar, or charged. In fact, the quadratic response function in eq. \eqref{eq:beta} contains only transition and fluctuation electric-dipole integrals, that, for $\pll \omu_{\alpha}; \omu_{\beta}, \omu_{\gamma} \prr_{\omega, \omega}$, at the new gauge-origin, read:
\begin{align}
\braket{0|\omu_{\alpha}|m}
&\rightarrow
\braket{0|\omu_{\alpha}|m}, \qquad \text{with } m \neq 0\\
\braket{m|\overline{\omu_{\alpha}}|n}
& \rightarrow
\braket{m|\overline{\omu_{\alpha}}|n}.
\end{align}
In fact, $-R_{\alpha}q_{\text{mol}}$ (the new term introducted by the gauge-origin shift) is not a quantum mechanical operator. At the same time, $\bbeta$ is origin independent also in the velocity representation, eq. \eqref{eq:beta_velocity}. Indeed, one can notice that only the electric-dipole moment in the the velocity formulation appears in the corresponding quadratic response function.
On the other hand, the mixed first hyperpolarizabilities $\aJ$, $\bJ$, $\aK$, and $\bK$ are origin-dependent. Consequently, for a shift of the gauge-origin, eq. \eqref{eq:O_new}, these mixed first hyperpolarizabilities change by:
\begin{align}
\bbeta & \rightarrow \bbeta^{\prime} = \bbeta,         \label{eq:beta_Op} \\
\aJ    & \rightarrow \aJ^{\prime} = \aJ + \Delta(\aJ), \label{eq:aJ_Op}   \\
\bJ    & \rightarrow \bJ^{\prime} = \bJ + \Delta(\bJ), \label{eq:bJ_Op}   \\
\aK    & \rightarrow \aK^{\prime} = \aK + \Delta(\aK), \label{eq:aK_Op}   \\
\bK    & \rightarrow \bK^{\prime} = \bK + \Delta(\bK). \label{eq:bK_Op}
\end{align}
where $\Delta(\aJ)$, $\Delta(\bJ)$, $\Delta(\aK)$, and $\Delta(\bK)$ are the gauge-origin shift tensors introduced by the change in the gauge-origin for $\aJ(\mathbf{O})$, $\bJ(\mathbf{O})$, $\aK(\mathbf{O})$, and $\bK(\mathbf{O})$. As said before, both in the length or in the velocity gauge, the gauge-origin shift tensor for $\bbeta$ is zero [i.e., $\Delta(\bbeta) = 0$]. The explicit form of these shifts depends on the formulation (length or velocity) of the response tensors. We examine them in the following.

\subsubsection{The length formulation} %
As already discussed in ref. \citenum{Bonvicini2023}, in the length formulation we have the following gauge-origin shifts:
\begin{align}
	\bbeta(\mathbf{O}^{\prime}) &\eqlen \Lambda_{\alpha\beta\gamma}
	\label{eq:beta_l_shift}\\
	\Delta(\aJ) &\eqlen + \frac{\iu}{2} |k^{\prime}| \epsilon_{\alpha b c} R_{b}
	\Lambda_{c\beta\gamma}
	\label{eq:aJ_l_shift}\\
	\Delta(\bJ) &\eqlen
	- \frac{\iu}{2}|k| \epsilon_{\gamma b c} R_{b} \Lambda_{\alpha\beta c}
	- \frac{\iu}{2}|k| \epsilon_{\beta b c} R_{b} \Lambda_{\alpha\gamma c}
	\label{eq:bJ_l_shift}\\
	\Delta(\aK) &\eqlen
	- \frac{1}{2} R_{\alpha} \Lambda_{\delta\beta\gamma}
	- \frac{1}{2} R_{\delta} \Lambda_{\alpha\beta\gamma}
	+ \frac{1}{3} R_{b}      \Lambda_{b\beta\gamma} \delta_{\alpha\delta}
	\label{eq:aK_l_shift}\\
	\begin{split}
	\Delta(\bK) &\eqlen
	- \frac{1}{2} R_{\gamma} \Lambda_{\alpha\beta\delta}
	- \frac{1}{2} R_{\delta} \Lambda_{\alpha\beta\gamma}
	+ \frac{1}{3} R_{b} \Lambda_{\alpha\beta b} \delta_{\gamma\delta}\\
& \quad
	- \frac{1}{2} R_{\beta} \Lambda_{\alpha\gamma\delta}
	- \frac{1}{2} R_{\delta} \Lambda_{\alpha\gamma\beta}
	+ \frac{1}{3} R_{b} \Lambda_{\alpha\gamma b} \delta_{\beta\delta}
	\label{eq:bK_l_shift}
	\end{split}
\end{align}
where, for compactness, we set $\Lambda_{\alpha\beta\gamma} \equiv \pll \omu_{\alpha}; \omu_{\beta}, \omu_{\gamma}\prr_{\omega, \omega}$, $|k'| = 2\omega/c_{0}$ and $|k| = \omega/c_{0}$. One can notice that all the shifts in eqs. \eqref{eq:aJ_l_shift}-\eqref{eq:bK_l_shift} contain $\bbeta$ written in the length formulation. In particular, as explained in ref. \citenum{Bonvicini2023}, the shifts for $\aJ$ and $\bJ$ can be written in terms of $\Lambda_{\alpha\beta\gamma}$ thanks to the hypervirial relations, i.e., they are valid only for exact wavefunctions or for approximated variational wavefunctions in the limit of a complete basis set.

\subsubsection{The velocity formulation} %
In the velocity formulation, the gauge-origin shifts on the first hyperpolarizabilities are:
%
\begin{align}
	\bbeta(\mathbf{O}^{\prime}) &\eqvel
	-\frac{\iu}{2\omega^{3}} \Pi_{\alpha\beta\gamma}
	\label{eq:beta_v_shift}\\
	\Delta(\aJ) &\eqvel
	+ \frac{1}{4\omega^{3}} |k^{\prime}| \epsilon_{\alpha b c} R_{b} \Pi_{c \beta\gamma}
	\label{eq:aJ_v_shift}\\
	\Delta(\bJ) &\eqvel
	- \frac{1}{4\omega^{3}}|k| \epsilon_{\gamma b c} R_{b} \Pi_{\alpha\beta c}
	- \frac{1}{4\omega^{3}}|k| \epsilon_{\beta b c} R_{b} \Pi_{\alpha\gamma c}
	\label{eq:bJ_v_shift}\\
	\Delta(\aK) &\eqvel
	+ \frac{\iu}{4\omega^{3}} R_{\alpha} \Pi_{\delta\beta\gamma}
	+ \frac{\iu}{4\omega^{3}} R_{\delta} \Pi_{\alpha\beta\gamma}
	- \frac{\iu}{6\omega^{3}} R_{b} \Pi_{b\beta\gamma} \delta_{\alpha\delta}
	\label{eq:aK_v_shift}\\
	\begin{split}
		\Delta(\bK) &\eqvel
		+ \frac{\iu}{4\omega^{3}} R_{\gamma} \Pi_{\alpha\beta\delta}
		+ \frac{\iu}{4\omega^{3}} R_{\delta} \Pi_{\alpha\beta\gamma}
		- \frac{\iu}{6\omega^{3}} R_{b} \Pi_{\alpha\beta b} \delta_{\gamma\delta}\\&\quad
		+ \frac{\iu}{4\omega^{3}} R_{\beta} \Pi_{\alpha\gamma\delta}
		+ \frac{\iu}{4\omega^{3}} R_{\delta} \Pi_{\alpha\gamma\beta}
		- \frac{\iu}{6\omega^{3}} R_{b} \Pi_{\alpha\gamma b} \delta_{\beta\delta}
		\label{eq:bK_v_shift}
	\end{split}
\end{align}
%
where, for compactness, we set $\Pi_{\alpha\beta\gamma} \equiv \pll \omu_{\alpha}^{p}; \omu_{\beta}^{p}, \omu_{\gamma}^{p} \prr_{\omega, \omega}$. As before, one can notice that all these tensor shifts contain $\bbeta$ written in the velocity formulation. There is a one-to-one correspondence between the length formulation, eqs. \eqref{eq:beta_l_shift}-\eqref{eq:bK_l_shift}, and the velocity formulation, eqs. \eqref{eq:beta_v_shift}-\eqref{eq:bK_v_shift}. In fact, by isolating the prefactor $(-1/(2\omega^{3})) = s$ in eqs. \eqref{eq:beta_v_shift}-\eqref{eq:bK_v_shift}, one obtains:
%
\begin{align}
\bbeta(\mathbf{O}^{\prime})	&\eqvel \iu s \;
\Pi_{\alpha\beta\gamma}
\label{eq:beta_v_shift_new},\\
\Delta(\aJ) 	&\eqvel \iu s \;
\bigg(
\frac{\iu}{2}
|k^{\prime}|
\epsilon_{\alpha b c}
R_{b}
\Pi_{c\beta\gamma}
\bigg)
\label{eq:aJ_v_shift_new}\\
\Delta(\bJ) 	&\eqvel \iu s \;
\bigg(
-\frac{\iu}{2}
|k|
\epsilon_{\gamma b c}
R_{b}
\Pi_{\alpha\beta c} 
-\frac{\iu}{2}
|k|
\epsilon_{\beta b c}
R_{b}
\Pi_{\alpha\gamma c}
\bigg)
\label{eq:bJ_v_shift_new}\\
\Delta(\aK) 	&\eqvel \iu s \;
\bigg(
-\frac{1}{2}
R_{\alpha}
\Pi_{\delta\beta\gamma}
-\frac{1}{2}
R_{\delta}
\Pi_{\alpha\beta\gamma}
+\frac{1}{3}
R_{c}
\Pi_{c\beta\gamma} \delta_{\alpha\delta}
\bigg)
\label{eq:aK_v_shift_new}\\
\begin{split}
\Delta(\bK) &\eqvel \iu s \;
\bigg(
-\frac{1}{2}
R_{\gamma}
\Pi_{\alpha\beta\delta}
-\frac{1}{2}
R_{\delta}
\Pi_{\alpha\beta\gamma}
+\frac{1}{3}
R_{c}
\Pi_{\alpha\beta c} \delta_{\gamma\delta}\\ &
-\frac{1}{2}
R_{\beta}
\Pi_{\alpha\gamma\delta}
-\frac{1}{2}
R_{\delta}
\Pi_{\alpha\gamma\beta}
+\frac{1}{3}
R_{c}
\Pi_{\alpha\gamma c} \delta_{\beta\delta}
\bigg)
\label{eq:bK_v_shift_new}
\end{split}
\end{align}
%
The rhs of eqs. \eqref{eq:beta_v_shift_new}-\eqref{eq:bK_v_shift_new} has the same structure as the rhs of eqs. \eqref{eq:beta_l_shift}-\eqref{eq:bK_l_shift}, i.e., there is a one-to-one correspondence between the two formulations. Indeed, these are related to each other by $\Lambda_{\alpha\beta\gamma} = \iu s \Pi_{\alpha\beta\gamma}$, which is valid thanks to the hypervirial relations.

\subsection{In the experimental observables} %
In general, any spectroscopic observable (e.g., intensity of the scattered light at the second harmonic for HRS-OA experiment) should be origin-independent. Consequently, the $a$, $b$, $c$, $d$, $f$, and $g$ terms (that contributes to the chiral and achiral intensity of the scattered light) should also be origin-independent.\cite{Bonvicini2023} For a shift on the gauge-origin, eq. \eqref{eq:O_new}, and by using eqs. \eqref{eq:beta_Op}-\eqref{eq:bK_Op} into the definitions of the $a$, $b$, $c$, $d$, $f$, and $g$ terms, eqs. \eqref{eq:a}-\eqref{eq:g}, the condition of origin independence for these terms reads:
\begin{align}
	a^{\prime} - a & = \langle \operatorname{Im}\{ \beta \Delta({}^{\beta}J)\} \rangle_{a} + |k|\langle \beta \Delta({}^{\beta}K) \rangle_{a}   = 0, \label{eq:a_shift}  \\
	b^{\prime} - b & = \langle \operatorname{Im}\{ \beta \Delta({}^{\alpha}J)\} \rangle_{b} + |k'|\langle \beta \Delta({}^{\alpha}K) \rangle_{b} = 0, \label{eq:b_shift} \\
	c^{\prime} - c & = \langle \operatorname{Im}\{ \beta \Delta({}^{\beta}J)\} \rangle_{c} + |k|\langle \beta \Delta({}^{\beta}K) \rangle_{c}   = 0, \label{eq:c_shift}  \\
	d^{\prime} - d & = |k'|\langle \beta \Delta({}^{\alpha}K) \rangle_{d} = 0,                                                  \label{eq:d_shift}                    \\
	f^{\prime} - f & = 0,                                                                                                   \label{eq:f_shift}                        \\
	g^{\prime} - g & = 0.																								    \label{eq:g_shift}
\end{align}
As explained before, the angular brackets in eqs. \eqref{eq:a_shift}-\eqref{eq:d_shift} indicates the molecular invariants (derived from the rotational averaging procedure) of the quantities inside them.
Because of the \textit{intrinsic} origin-independence of $\bbeta$ in both the length and velocity formulations, the achiral $f$ and $g$ terms [eqs. \eqref{eq:f}-\eqref{eq:g}] are \textit{automatically} origin-independent and no further discussion is required for them. Therefore. we will from this point concentrate our discussion on the chiral $a$, $b$, $c$, and $d$ terms, only. For $a$, $b$, and $c$ the condition of origin-independence is guaranteed as a combined effect of two terms. On the other hand, the condition of origin-independence for $d$ is guaranteed by $\langle \beta \Delta({}^{\alpha}K) \rangle_{d}$ \textit{alone} ($|k'|$ is clearly nonzero). This difference is important for the discussion that concerns the length formulation, as detailed in the next paragraph.

\subsubsection{Length}
Is has been proved in ref. \citenum{Bonvicini2023} that, by inserting the gauge-origin shifts expressed in the length formulation [eqs. \eqref{eq:beta_l_shift}-\eqref{eq:bK_l_shift}] into the conditions of origin-independence of the theory, eqs. \eqref{eq:a_shift}-\eqref{eq:d_shift}, and by expanding all the molecular invariants, the chiral terms $a$, $b$, $c$, and $d$ are origin-independent. In particular, $a$, $b$, $c$ are origin-independent as long as the hypervirial relations are fulfilled (exact wavefunctions or approximated wavefunctions in the limit of a complete basis set). The chiral term $d$ is origin-independent \textit{by construction} because satisfying eq. \eqref{eq:d_shift} does not require the use of the hypervirial relations (in fact, it does not contain any $\Delta({}^{\alpha}J)$ or $\Delta({}^{\beta}J)$ term).

\subsubsection{Velocity}
Because of the one-to-one correspondence between the gauge-origin shifts in the length [eqs. \eqref{eq:beta_l_shift}-\eqref{eq:bK_l_shift}] and the velocity [eqs. \eqref{eq:beta_v_shift_new}-\eqref{eq:bK_v_shift_new}] formulations, the velocity one is also origin-independent. In fact, symbolically speaking, the $\Lambda_{\alpha\beta\gamma}$ and $\Pi_{\alpha\beta\gamma}$ tensors play the same role once summed in the molecular invariants present in eqs. \eqref{eq:a_shift}-\eqref{eq:d_shift}. This result is not completely surprising because we simply found an alternative way to express $\bbeta$, $\aJ$, $\bJ$, $\aK$, and $\bK$. One formulation or the other should be irrelevant for eqs. \eqref{eq:a}-\eqref{eq:g}. On the other hand, the strength of this work lies on the fact that the velocity formulation of the pure and mixed first hyperpolarizabilities provides a theory which is origin-independent by construction.
Consequently, it is origin independent for calculation with exact and, more importantly, with approximated (variational and not) wavefunctions. The length formulation is origin-independent only for exact wavefunctions because it is necessary to invoke the hypervirial relations in order to obtain eqs. \eqref{eq:aJ_l_shift} and \eqref{eq:bJ_l_shift} in that form that contains $\bbeta$ written in the length formulation.

\section{Computational details} %
The origin-independence of the velocity formulation has been tested for a prototypical chiral molecule: $R$-methyloxirane, whose geometry has been optimized in a previous work.\cite{Bonvicini2025a}
The first hyperpolarizabilities have been computed with the DALTON quantum chemistry software\cite{Aidas2014, Olsen2020a} using time-dependent Hartree-Fock (TD-HF) level and different basis sets containing different sets of diffuse functions: cc-pVXZ, aug-cc-pVXZ, and d-aug-cc-pVXZ, with X=D, T, and Q quality. The results are discussed in the next Section.

\section{Illustrations} %
\begin{figure*}[t!]
	\includegraphics[width=1.0\linewidth]{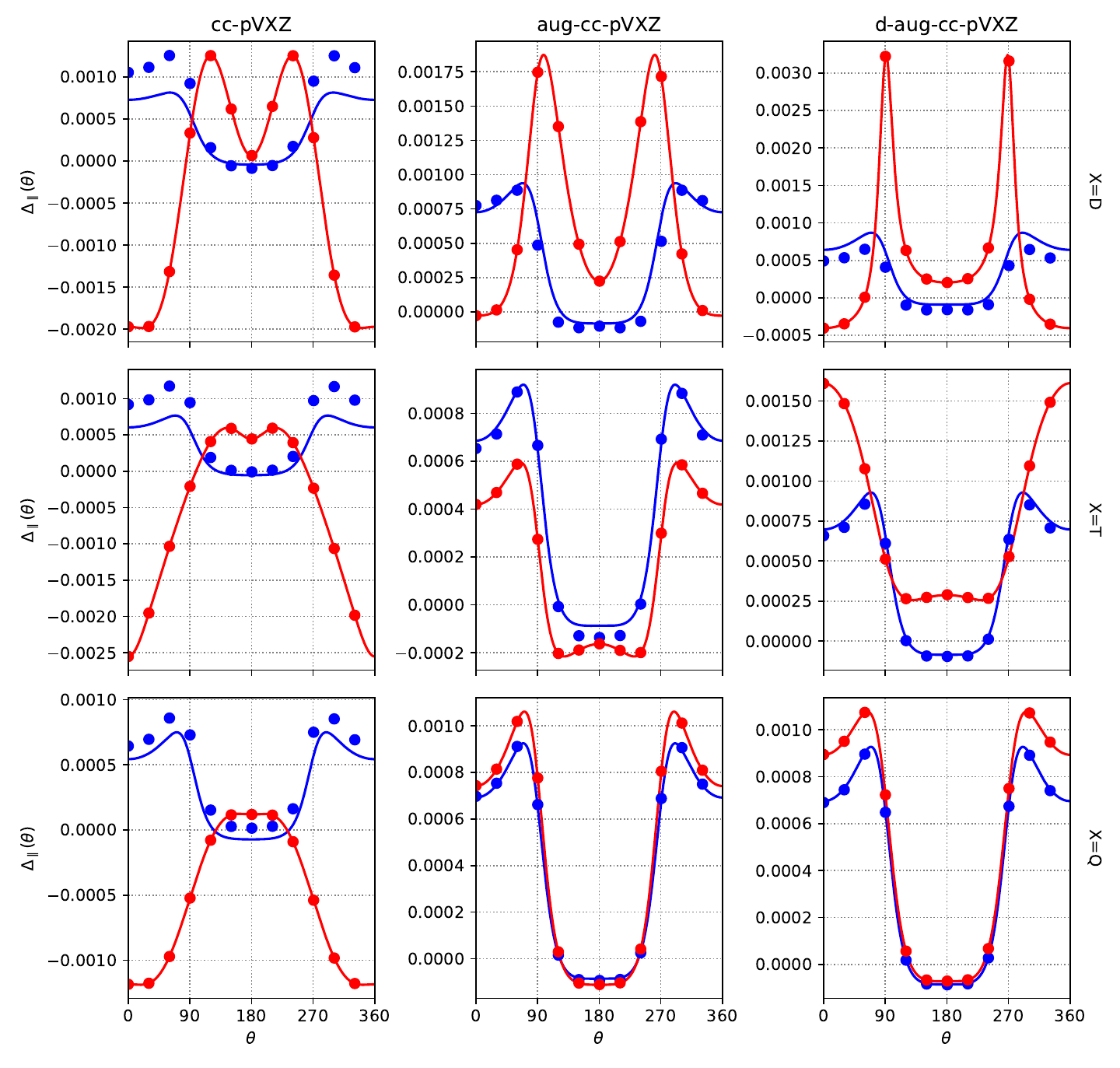}
	\caption{\label{fig:Figure_1}Profiles of the differential scattering ratios $\Delta_{\parallel}(\theta)$ as a function of the scattering angle $\theta$, in degrees, computed at the TD-HF level with different basis sets. The lines refers to the gauge origin $\mathbf{O}$ (the COM of the molecule is at the origin of the molecular frame), while the circles refer to the new gauge origin $\mathbf{O}^{\prime}$ where the COM of the molecule has been diplaced by $(10,10,10) \, \si{\angstrom}$. The blue and red colors (for lines/points) refers to the length and velocity formulations, respectively.}
\end{figure*}

\begin{figure*}[t!]
	\includegraphics[width=1.0\linewidth]{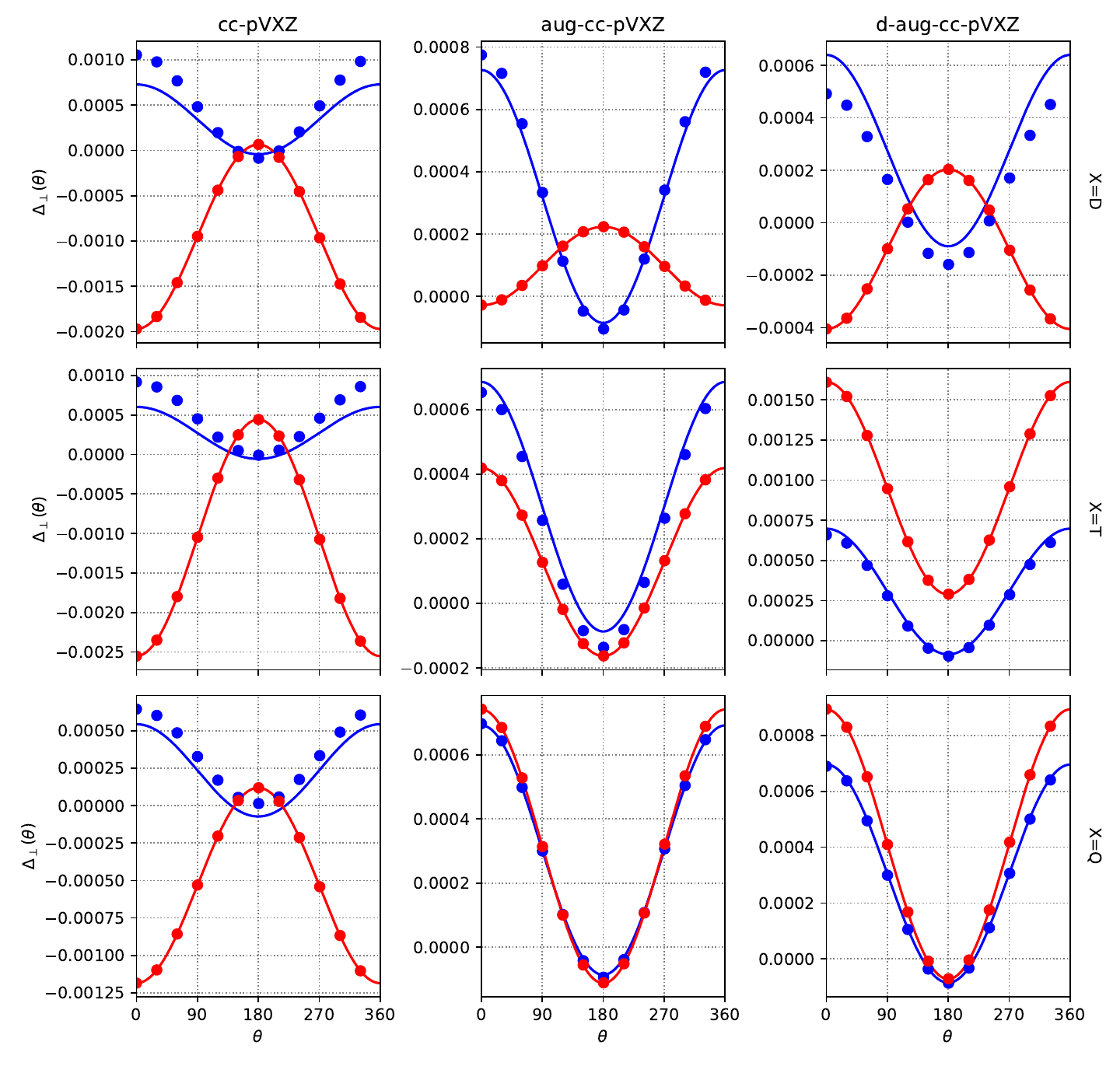}
	\caption{\label{fig:Figure_2}Profiles of the differential scattering ratios $\Delta_{\perp}(\theta)$ as a function of the scattering angle $\theta$, in degrees, computed at the TD-HF level with different basis sets. The lines refers to the gauge origin $\mathbf{O}$ (the COM of the molecule is at the origin of the molecular frame), while the circles refer to the new gauge origin $\mathbf{O}^{\prime}$ where the COM of the molecule has been diplaced by $(10,10,10) \, \si{\angstrom}$. The blue and red colors (for lines/points) refers to the length and velocity formulations, respectively.}
\end{figure*}

The centre-of-mass (COM) of $R$-methyloxirane has been placed at two different points in space: i) at the origin of the molecular frame [i.e., at $(0,0,0)\,\si{\angstrom}$], and ii) then displaced by a vector $\textbf{R}_{\text{mol}} = (10,10,10) \, \si{\angstrom}$. Consequently, there are two gauge-origins: $\mathbf{O} = (0, 0, 0) \, \si{\angstrom}$ and $\mathbf{O}^{\prime} = -(10, 10, 10) \, \si{\angstrom}$.
The results are presented in Table \ref{tab:origin-independence} and in Figures \ref{fig:Figure_1} and \ref{fig:Figure_2}. The error introduced by changing the gauge origin from $\mathbf{O}$ to $\mathbf{O}^{\prime}$ has been computed as:
\begin{align}
\Delta_{\mathbf{O} \rightarrow \mathbf{O}^{\prime}}
=
\frac{|t^{\prime} - t|}{|t|} \times 100, \qquad\text{with } t \in \{a, b, c, d, f, g\}
\label{eq:gauge_error}
\end{align}
where the prime is associated with the new gauge origin $\mathbf{O}^{\prime}$.
\begin{table*}[t!]
	\caption{\label{tab:origin-independence}Comparison between the length vs. the velocity formulation of HRS-OA for two different gauge origins, $\mathbf{O}$ and $\mathbf{O}^{\prime}$, for the chiral $a$, $b$, $c$, and $d$, and achiral $f$ and $g$ terms. The error introduced by the change in the gauge origin, $\Delta_{\mathbf{O}^{\prime},\mathbf{O}}$, is given in percent [eq. \eqref{eq:gauge_error}].}
	\begin{ruledtabular}
\begin{tabular}{r|r|rr|rr|rr}
	                               &      & \multicolumn{2}{c|}{$\mathbf{O}$} & \multicolumn{2}{c|}{$\mathbf{O}^{\prime}$} & \multicolumn{2}{c}{$\Delta_{\mathbf{O} \rightarrow \mathbf{O}^{\prime}}$} \\ \hline
	                     Basis set & Term &     Length &             Velocity &     Length &                      Velocity & Length &                 Velocity \\ \hline\hline
	      \multirow{6}{*}{cc-pVDZ} &  $a$ &     18.311 &             2933.839 &     30.448 &                      2933.719 & 66.278 &                    0.004 \\
	                               &  $b$ &     34.673 &           -29121.155 &     52.397 &                    -29124.966 & 51.117 &                    0.013 \\
	                               &  $c$ &     14.589 &           -13031.677 &     16.034 &                    -13032.408 &  9.902 &                    0.006 \\
	                               &  $d$ &      2.345 &            18334.474 &      2.345 &                     18334.083 &  0.001 &                    0.002 \\
	                               &  $f$ &  32766.965 &          9174084.539 &  32766.965 &                   9174084.539 &  0.000 &                    0.000 \\
	                               &  $g$ &  63309.517 &          1427872.087 &  63309.517 &                   1427872.087 &  0.000 &                    0.000 \\ \hline
	      \multirow{6}{*}{cc-pVTZ} &  $a$ &     13.004 &            -1670.437 &     21.888 &                     -1670.295 & 68.316 &                    0.009 \\
	                               &  $b$ &     25.487 &           -11107.915 &     36.484 &                    -11108.127 & 43.146 &                    0.002 \\
	                               &  $c$ &      9.366 &            -1760.171 &     15.294 &                     -1759.867 & 63.287 &                    0.017 \\
	                               &  $d$ &      1.496 &             6235.828 &      1.496 &                      6235.981 &  0.005 &                    0.002 \\
	                               &  $f$ &  23021.271 &          7737996.161 &  23021.271 &                   7737996.161 &  0.000 &                    0.000 \\
	                               &  $g$ &  58878.349 &         -4484576.933 &  58878.349 &                  -4484576.933 &  0.000 &                    0.000 \\ \hline
	      \multirow{6}{*}{cc-pVQZ} &  $a$ &      9.701 &            -1709.110 &     12.773 &                     -1709.181 & 31.668 &                    0.004 \\
	                               &  $b$ &     20.333 &            -3075.643 &     20.819 &                     -3075.392 &  2.388 &                    0.008 \\
	                               &  $c$ &      6.047 &              473.516 &      9.233 &                       474.262 & 52.695 &                    0.157 \\
	                               &  $d$ &      0.262 &             1564.556 &      0.262 &                      1564.469 &  0.004 &                    0.006 \\
	                               &  $f$ &  17381.929 &          3244131.540 &  17381.929 &                   3244131.540 &  0.000 &                    0.000 \\
	                               &  $g$ &  49468.371 &          -926561.417 &  49468.371 &                   -926561.417 &  0.000 &                    0.000 \\ \hline
	  \multirow{6}{*}{aug-cc-pVDZ} &  $a$ &     7.8665 &             296.6296 &     5.8400 &                      296.7541 & 25.761 &                    0.042 \\
	                               &  $b$ &    17.3579 &            -295.5966 &    18.8206 &                     -295.2897 &  8.427 &                    0.104 \\
	                               &  $c$ &     5.9224 &            -244.9896 &     8.5782 &                     -244.8758 & 44.844 &                    0.046 \\
	                               &  $d$ &     0.0834 &             229.0047 &     0.0834 &                      228.7395 &  0.060 &                    0.116 \\
	                               &  $f$ & 11840.6809 &          170625.9884 & 11840.6809 &                   170625.9884 &  0.000 &                    0.000 \\
	                               &  $g$ & 31161.7678 &          358515.1726 & 31161.7678 &                   358515.1726 &  0.000 &                    0.000 \\ \hline
	  \multirow{6}{*}{aug-cc-pVTZ} &  $a$ &     7.1184 &               5.8918 &     7.3122 &                        5.9239 &  2.723 &                    0.544 \\
	                               &  $b$ &    16.1287 &              30.9164 &    16.4699 &                       30.9148 &  2.116 &                    0.005 \\
	                               &  $c$ &     5.1225 &               5.0941 &     3.2614 &                        5.0973 & 36.332 &                    0.061 \\
	                               &  $d$ &    -0.3327 &              -5.9321 &    -0.3326 &                       -5.9414 &  0.006 &                    0.157 \\
	                               &  $f$ & 10871.0784 &           21143.5191 & 10871.0784 &                    21143.5191 &  0.000 &                    0.000 \\
	                               &  $g$ & 30010.4913 &           64655.6808 & 30010.4913 &                    64655.6808 &  0.000 &                    0.000 \\ \hline
	  \multirow{6}{*}{aug-cc-pVQZ} &  $a$ &     7.0009 &               9.1681 &     7.1102 &                        9.1658 &  1.561 &                    0.025 \\
	                               &  $b$ &    15.8544 &              19.5304 &    16.1114 &                       19.5406 &  1.621 &                    0.052 \\
	                               &  $c$ &     5.1005 &               4.4053 &     4.9463 &                        4.4611 &  3.022 &                    1.266 \\
	                               &  $d$ &    -0.3033 &              -1.1073 &    -0.3033 &                       -1.1100 &  0.003 &                    0.243 \\
	                               &  $f$ & 10640.8260 &           11703.3946 & 10640.8260 &                    11703.3946 &  0.000 &                    0.000 \\
	                               &  $g$ & 29322.6248 &           31447.6761 & 29322.6248 &                    31447.6761 &  0.000 &                    0.000 \\ \hline
	\multirow{6}{*}{d-aug-cc-pVDZ} &  $a$ &     6.5348 &             195.2997 &     4.3696 &                      195.3117 & 33.133 &                    0.006 \\
	                               &  $b$ &    14.9762 &            -177.8708 &    13.4017 &                     -178.2722 & 10.514 &                    0.226 \\
	                               &  $c$ &     4.4874 &            -320.1096 &     2.3033 &                     -320.0563 & 48.672 &                    0.017 \\
	                               &  $d$ &    -0.3742 &            -200.1672 &    -0.3741 &                     -199.9380 &  0.021 &                    0.115 \\
	                               &  $f$ & 10552.3857 &           60830.2783 & 10552.3857 &                    60830.2783 &  0.000 &                    0.000 \\
	                               &  $g$ & 29469.3577 &         1181629.0483 & 29469.3577 &                  1181629.0483 &  0.000 &                    0.000 \\ \hline
	\multirow{6}{*}{d-aug-cc-pVTZ} &  $a$ &     7.0291 &               5.4521 &     6.5599 &                        5.4563 &  6.675 &                    0.079 \\
	                               &  $b$ &    15.8669 &               9.5551 &    15.3002 &                        9.5246 &  3.572 &                    0.320 \\
	                               &  $c$ &     5.1921 &              11.5922 &     4.6713 &                       11.5911 & 10.030 &                    0.010 \\
	                               &  $d$ &    -0.2446 &               2.3205 &    -0.2446 &                        2.3202 &  0.008 &                    0.015 \\
	                               &  $f$ & 10636.0120 &           10574.5940 & 10636.0120 &                    10574.5940 &  0.000 &                    0.000 \\
	                               &  $g$ & 29277.2203 &            7377.1977 & 29277.2203 &                     7377.1977 &  0.000 &                    0.000 \\ \hline
	\multirow{6}{*}{d-aug-cc-pVQZ} &  $a$ &     7.0008 &               7.7619 &     6.9432 &                        7.7595 &  0.823 &                    0.030 \\
	                               &  $b$ &    15.8068 &              16.6945 &    15.7208 &                       16.6995 &  0.544 &                    0.030 \\
	                               &  $c$ &     5.1627 &               5.9770 &     5.0721 &                        6.0163 &  1.754 &                    0.656 \\
	                               &  $d$ &    -0.2464 &              -0.5473 &    -0.2464 &                       -0.5468 &  0.008 &                    0.095 \\
	                               &  $f$ & 10608.6467 &           10637.8592 & 10608.6467 &                    10637.8592 &  0.000 &                    0.000 \\
	                               &  $g$ & 29240.9853 &           22822.4835 & 29240.9853 &                    22822.4835 &  0.000 &                    0.000
\end{tabular}
	\end{ruledtabular}
\end{table*}
From Table \ref{tab:origin-independence} it is evident that the length formulation is origin-dependent, while the velocity formulation is origin-independent for any basis set size. In fact, the chiral $a$, $b$, and $c$ terms are origin-dependent in the length formulation (Table \ref{tab:origin-independence}). Moreover, as expected, the origin-dependence of the length formulation is attenuated by increasing the basis set size (Table \ref{tab:origin-independence}).
It is important to notice that the chiral $d$ term is origin-independent in both the length and velocity formulation (Table \ref{tab:origin-independence}). In fact, the $d$ term, eq. \eqref{eq:d}, contains molecular invariants made by the interference of $\bbeta$ (which is intrinsically origin-independent) with the origin-dependent $\aK$ first hyperpolarizability, whose gauge-origin shifts, $\langle \beta \Delta({}^{\alpha}K) \rangle_{d}$, sum up to zero by construction (i.e., they do not require the hypervirial relations to satisfy the origin-independence) in both the length and the velocity formulations, see eq. \eqref{eq:d_shift} and ref. \citenum{Bonvicini2023}.
Finally, as expected, the achiral $f$ and $g$ terms, which contain the interference of $\bbeta$ with itself, are origin-independent in both formulations because the pure electric-dipole first hyperpolarizability is \textit{intrinsically} an origin-independent molecular response for exact as well approximated wavefunctions and for any molecular system (neutral, polar, or charged).
Concerning the profiles of the differential scattering ratios, $\Delta_{\parallel,\perp}(\theta)$, as a function of the scattering angle $\theta$ (Figures \ref{fig:Figure_1} and \ref{fig:Figure_2}), one can notice that for the velocity formulation, the red lines (origin $\mathbf{O}$) and points (origin $\mathbf{O}^{\prime}$) are perfectly superimposed for all the basis sets, thus showing the origin-independence of the velocity formulation in the central quantity in a HRS-OA experiment. For small basis sets without diffuse functions, the length and velocity formulation provide very different profiles in both quantitative and qualitative aspects. However, as expected, by introducing diffuse functions and increasing the angular momentum in the basis set, the length and the velocity profiles are more and more similar both quantitatively and qualitatively. However, a perfect match between the length and the velocity formulations is not achieved for any basis set considered here.
Moreover, it is interesting to notice a rather surprising result: the profiles for both $\Delta_{\parallel}(\theta)$ and $\Delta_{\perp}(\theta)$ computed with the length and the velocity formulations are qualitatively and quantitatively more similar in the aug-cc-pVXZ basis compared with the d-aug-cc-pVXZ basis, for X=D,T,Q (Figures \ref{fig:Figure_1} and \ref{fig:Figure_2}).
Finally, it is quite evident from Table \ref{tab:origin-independence}, Figure \ref{fig:Figure_1}, and Figure \ref{fig:Figure_2} that the results obtained with the velocity formulation are more influenced by the basis set size compared to the length ones. However, this is a known issue that affects not only the velocity formulation discussed here but other velocity formulations like those that have been presented in the Introduction.

\section{Conclusions} %
In this work, two velocity formulations of HRS-OA have been presented: the velocity, eqs. \eqref{eq:beta_velocity}-\eqref{eq:bK_velocity}, and the full velocity, eqs. \eqref{eq:aJ_velocity_full}-\eqref{eq:bK_velocity_full}, formulations of the pure and mixed first hyperpolarizabilities which are necessary for the simulation of the HRS-OA observables. It has been showed that both of them are origin-independent by construction (i.e., the origin-independence property of these formulations do not require the use of the hypervirial relations, which are instead required in the length formulation of HRS-OA). Consequently, calculations with the velocity formulation of HRS-OA will always provide origin-independent results also for approximated (variational) wavefunctions.
Numerical illustrations based on quantum chemistry calculations confirmed that the velocity formulation of HRS-OA remains origin-independent for any basis set size, while the length formulation remains origin-dependent and it reaches origin-independence only in the limit of a completes basis set. This origin-dependence of the length formulation can be reduced by increasing the basis set size, especially by including both diffuse and higher angular momentum functions.
In future works, we consider the derivation of the velocity formulation for the third-harmonic scattering optical activity (THS-OA) spectroscopy.\cite{Bonvicini2023a}

\section{Acknowledgments} %
A.B. is a postdoctoral researcher of the Fonds de la Recherche Scientifique—FNRS (F.R.S.-FNRS).
A.B. and B.C. acknowledge the F.R.S.-FNRS (Nonlinear chiroptics projects) for the grant nos. T.0025.22 and T.0052.24.
The calculations were performed on the computers of the “Consortium des Équipements de Calcul Intensif (CÉCI)” (http://www.ceci-hpc.be), including those of the “UNamur Technological Platform of High-Performance Computing (PTCI)” (http://www.ptci.unamur.be) and those of the Tier-1 supercomputer of the Fédération Wallonie-Bruxelles, for which the authors gratefully acknowledge the financial support from FNRS-FRFC, the Walloon Region, and the University of Namur (Conventions no. 2.5020.11, U.G006.15, U.G018.19, U.G011.22, RW1610468, RW/GEQ2016, RW1910247, RW2110213, and RW2210148).

\section*{SUPPLEMENTARY MATERIAL} %
The derivation of the commutators in eqs. \eqref{eq:comm_mu_H}-\eqref{eq:comm_Q_imup} and the velocity formulation in eqs. \eqref{eq:beta_velocity}-\eqref{eq:bK_velocity} are provided in the Supporting Information file.

\section*{DATA AVAILABILITY} %
The data that support the findings of this study are available
within the article and its supplementary material.

\bibliography{library.bib}

\clearpage
\begin{widetext}
\appendix
\input{SI.tex}
\end{widetext}

\end{document}

%% file: SI.tex
\setcounter{section}{0}
\renewcommand{\thesection}{S\arabic{section}}
\setcounter{page}{1}
\renewcommand{\thepage}{S\arabic{page}}
\setcounter{table}{0}
\renewcommand{\thetable}{S\arabic{table}}
\setcounter{equation}{0}
\renewcommand{\theequation}{S\arabic{equation}}
\setcounter{figure}{0}
\renewcommand{\thefigure}{S\arabic{figure}}


\section{Commutators} %
\label{sec:commutators}
All the commutators obtained in the following subsections are derived assuming electronic wavefunctions in the frozen nuclei approximation.
Consequently, these commutators are evaluated for the electrons only, and the generic indices ($i$ and $j$) in the sums over the particles are meant to be sums over the $N_{e}$ electrons of the molecule.
Finally, atomic units (i.e., $\hbar = e = m_{e} = 1$) are employed in the last equation of each derivation in order to provide the simplest and shortest expression.

\subsection{$[\omu_{\alpha}, \hat{H}_{\normalfont{\text{mol}}}]$} %
\label{sec:comm_mu_H}

\begin{align}
	[\omu_{\alpha}, \hat{H}_{\text{mol}}]
	&=
	\left[
	\sum_{i} q_{i} \hat{r}_{i,\alpha}
	,
	\sum_{j} \hat{V}(\mathbf{r}_{j}) +
	\frac{\hat{p}_{j,\beta}\hat{p}_{j,\beta}}{2m_{j}}
	\right]\\
	&=
	\frac{1}{2}
	\sum_{i}\sum_{j}
	\frac{q_{i}}{m_{j}}
	\left[
	\hat{r}_{i,\alpha}
	,
	\hat{p}_{j,\beta}\hat{p}_{j,\beta}
	\right] \label{eq:comm_mu_H_step_1a} \\
	&=
	\frac{1}{2}
	\sum_{i}\sum_{j}
	\frac{q_{i}}{m_{j}}
	\left(
	\left[
	\hat{r}_{i,\alpha}
	,
	\hat{p}_{j,\beta}
	\right]
	\hat{p}_{j,\beta}
	+
	\hat{p}_{j,\beta}
	\left[
	\hat{r}_{i,\alpha}
	,
	\hat{p}_{j,\beta}
	\right]
	\right) \label{eq:comm_mu_H_step_1b} \\
	&=
	\frac{1}{2}
	\sum_{i}\sum_{j}
	\frac{q_{i}}{m_{j}}
	2
	\iu
	\hbar
	\delta_{ij}
	\delta_{\alpha\beta}
	\hat{p}_{j,\beta}
	\\
	&=
	\iu
	\hbar
	\sum_{i}
	\frac{q_{i}}{m_{i}}
	\hat{p}_{i,\alpha}
	\\
	&=
	\iu
	\omu_{\alpha}^{p}
\end{align}
Ongoing from eq. \eqref{eq:comm_mu_H_step_1a} to eq. \eqref{eq:comm_mu_H_step_1b} we used the following identity:
\begin{equation}
	[A, BC] = [A, B] C + B[A, C] \label{eq:comm_[A,BC]}.
\end{equation}
While, in the last step, atomic units ($\hbar = 1$) and the definition of the electric-dipole moment operator in the velocity formulation, $\omu_{\alpha}^{p}$ [eq. \eqref{eq:e_dip_p}] have been used.

\clearpage
\subsection{$[\omu_{\alpha}, \omu_{\beta}]$} %
\label{sec:comm_mu_mu}

\begin{align}
	[\omu_{\alpha}, \omu_{\beta}]
	&=
	\left[
	\sum_{i}
	q_{i}
	\hat{r}_{i,\alpha}
	,
	\sum_{j}
	q_{j}
	\hat{r}_{j,\beta}
	\right]\\
	&=
	\sum_{i}\sum_{j}
	q_{i}q_{j}
	[
	\hat{r}_{i,\alpha},
	\hat{r}_{j,\beta}
	]\\
	&= 0
\end{align}
This commutator is equal to zero because two position operators always commute i.e., $[\hat{r}_{i,\alpha}, \hat{r}_{j,\beta}] = 0$.

\clearpage
\subsection{$[\omu_{\alpha}, \oqu_{\beta\gamma}]$} %
\label{sec:comm_mu_qu}

\begin{align}
	[\omu_{\alpha}, \oqu_{\beta\gamma}]
	&=
	\left[
	\sum_{i} q_{i} \hat{r}_{i,\alpha}
	,
	\frac{1}{2}\sum_{j} q_{j} (\hat{r}_{j,\beta} \hat{r}_{j,\gamma}
	-\frac{1}{3} \hat{r}_{j,\delta} \hat{r}_{j,\delta} \delta_{\beta\gamma})
	\right]\\
	&=
	\frac{1}{2}
	\sum_{i}
	\sum_{j}
	q_{i} q_{j}
	\left[
	\hat{r}_{i,\alpha}
	,
	\hat{r}_{j,\beta} \hat{r}_{j,\gamma}
	-\frac{1}{3} \hat{r}_{j,\delta} \hat{r}_{j,\delta} \delta_{\beta\gamma}
	\right]\\
	&=
	\frac{1}{2}
	\sum_{i}
	\sum_{j}
	q_{i} q_{j}
	\left(
	\left[
	\hat{r}_{i,\alpha}
	,
	\hat{r}_{j,\beta} \hat{r}_{j,\gamma}
	\right]
	-\frac{1}{3}\delta_{\beta\gamma}
	\left[
	\hat{r}_{i,\alpha}
	,
	\hat{r}_{j,\delta} \hat{r}_{j,\delta}
	\right]
	\right)\\
	&=
	\frac{1}{2}
	\sum_{i}
	\sum_{j}
	q_{i} q_{j}
	\left(
	\left[
	\hat{r}_{i,\alpha}
	,
	\hat{r}_{j,\beta}
	\right]
	\hat{r}_{j,\gamma}
	+
	\hat{r}_{j,\beta}
	\left[
	\hat{r}_{i,\alpha}
	,
	\hat{r}_{j,\gamma}
	\right]
	-\frac{1}{3}\delta_{\beta\gamma}
	\left[
	\hat{r}_{i,\alpha}
	,
	\hat{r}_{j,\delta}
	\right]
	\hat{r}_{j,\delta}
	-\frac{1}{3}\delta_{\beta\gamma}
	\hat{r}_{j,\delta}
	\left[
	\hat{r}_{i,\alpha}
	,
	\hat{r}_{j,\delta}
	\right]
	\right) \label{eq:comm_mu_qu_expansion}\\
	&= 0
\end{align}
For the same reason as before, this commutator is also equal to zero. In fact, by using the commutator identity for $[A, BC]$, eq. \eqref{eq:comm_[A,BC]}, one can notice that only commutators between position operators are present in eq. \eqref{eq:comm_mu_qu_expansion}.

\clearpage
\subsection{$[\omu_{\alpha}, \omu_{\beta}^{p}]$} %
\label{sec:comm_mu_mup}

\begin{align}
	[\omu_{\alpha}, \omu_{\beta}^{p}]
	&=
	\left[
	\sum_{i}
	q_{i}
	\hat{r}_{i,\alpha}
	,
	\sum_{j}
	\frac{q_{j}}{m_{j}}
	\hat{p}_{j,\beta}
	\right]\\
	&=
	\sum_{i}
	\sum_{j}
	\frac{q_{i}q_{j}}{m_{j}}
	\left[
	\hat{r}_{i,\alpha}
	,
	\hat{p}_{j,\beta}
	\right]\\
	&=
	\iu
	\hbar
	\sum_{i}
	\sum_{j}
	\frac{q_{i}q_{j}}{m_{j}}
	\delta_{ij}
	\delta_{\alpha\beta}
	\\
	&=
	\iu
	\hbar
	\sum_{i}
	\frac{q_{i}q_{i}}{m_{i}}
	\delta_{\alpha\beta}
	\\
	&=
	\iu
	N_{e}
	\delta_{\alpha\beta}
\end{align}
Where atomic units have been employed ($\hbar = -q_{i} = m_{i} = e = 1$) and $N_{e}$ is the total number of electrons in the system (molecule).

\clearpage
\subsection{$[\omu_{\alpha}, \oma_{\beta}]$} %
\label{sec:comm_mu_m}                        %

%
\begin{align}
	[\omu_{\alpha}, \oma_{\beta}]
	&=
	\left[
	\sum_{i} q_{i} \hat{r}_{i,\alpha}, \frac{1}{2c_{0}} \sum_{j} \frac{q_{j}}{m_{j}} \epsilon_{\beta c d} \hat{r}_{j, c} \hat{p}_{j, d}
	\right]\\
	&=
	\frac{1}{2c_{0}}
	\sum_{i}\sum_{j} \frac{q_{i}q_{j}}{m_{j}}
	\epsilon_{\beta c d}
	\left[
	\hat{r}_{i,\alpha},
	\hat{r}_{j, c} \hat{p}_{j, d}
	\right] \label{eq:comm_mu_m_step_1a}\\
	&=
	\frac{1}{2c_{0}}
	\sum_{i}\sum_{j} \frac{q_{i}q_{j}}{m_{j}}
	\epsilon_{\beta c d}
	\left(
	\left[
	\hat{r}_{i,\alpha},
	\hat{r}_{j, c}
	\right]
	\hat{p}_{j, d}
	+
	\hat{r}_{j, c}
	\left[
	\hat{r}_{i,\alpha},
	\hat{p}_{j, d}
	\right]
	\right) \label{eq:comm_mu_m_step_1b}\\
	&=
	\frac{1}{2c_{0}}
	\sum_{i}\sum_{j} \frac{q_{i}q_{j}}{m_{j}}
	\epsilon_{\beta c d}
	\hat{r}_{j, c}
	\left[
	\hat{r}_{i,\alpha},
	\hat{p}_{j, d}
	\right]\\
	&=
	\frac{1}{2c_{0}}
	\sum_{i}\sum_{j} \frac{q_{i}q_{j}}{m_{j}}
	\epsilon_{\beta c d}
	\hat{r}_{j, c}
	\iu
	\hbar
	\delta_{ij}
	\delta_{\alpha d}\\
	&=
	\frac{\iu\hbar}{2c_{0}}
	\sum_{i} \frac{q_{i}q_{i}}{m_{i}}
	\epsilon_{\beta c d}
	\hat{r}_{i, c}
	\delta_{\alpha d}\\
	&=
	\frac{\iu\hbar}{2c_{0}}
	\epsilon_{\beta c\alpha}
	\sum_{i} \frac{q_{i}q_{i}}{m_{i}}
	\hat{r}_{i, c}\\
	&=
	\frac{\iu\hbar}{2c_{0}}
	\epsilon_{\alpha\beta c}
	\sum_{i} \frac{q_{i}q_{i}}{m_{i}}
	\hat{r}_{i, c} \label{eq:comm_mu_m_step_2a}\\
	&=
	-
	\frac{\iu}{2c_{0}}
	\epsilon_{\alpha\beta c}
	\omu_{ c} \label{eq:comm_mu_m_step_2b}
\end{align}
Ongoing from eq. \eqref{eq:comm_mu_m_step_1a} to eq. \eqref{eq:comm_mu_m_step_1b} we used the commutator identity for $[A, BC]$ given in eq. \eqref{eq:comm_[A,BC]}.
While from eq. \eqref{eq:comm_mu_m_step_2a} to eq. \eqref{eq:comm_mu_m_step_2b} we used atomic units ($\hbar = m_{i} = e = 1$ and $q_{i} = -e = -1$) and electronic wavefunctions, together with the definition of the electric-dipole moment operator [eq. \eqref{eq:e_dip}].

\clearpage
\subsection{$[\oqu_{\alpha\beta}, \hat{H}_{\normalfont\text{mol}}]$} %
\label{sec:comm_Q_H}                                                 %
In order to obtain the velocity formulation for the mixed electric-dipole/electric-quadrupole first hyperpolarizabilities, $\aK$ and $\bK$, it is necessary to consider the commutator between the traceless electric-quadrupole moment operator, eq. \eqref{eq:e_quad}, and the molecular electronic Hamiltonian, $\hat{H}_{\text{mol}}$, which reads:
\begin{align}
	[\oqu_{\alpha\beta}, \hat{H}_{\text{mol}}] =
	\frac{1}{2} \left[\sum_{i} q_{i} \hat{r}_{i,\alpha} \hat{r}_{i,\beta}, \hat{H}_{\text{mol}}\right]
	-
	\frac{1}{6} \left[ \sum_{i} q_{i} \hat{r}_{i,\gamma} \hat{r}_{i,\gamma} \delta_{\alpha\beta}, \hat{H}_{\text{mol}} \right]
	\label{eq:Q_H_commutator}
\end{align}
Let consider the first term in the rhs of eq. \eqref{eq:Q_H_commutator}:
\begin{align}
	\frac{1}{2} \left[\sum_{i} q_{i} \hat{r}_{i,\alpha} \hat{r}_{i,\beta}, \hat{H}_{\text{mol}}\right]
	&=
	\frac{1}{2} \left[\sum_{i} q_{i} \hat{r}_{i,\alpha} \hat{r}_{i,\beta}, \sum_{j} \hat{V}(\mathbf{r}_{j}) + \frac{\hat{p}_{j,\gamma} \hat{p}_{j,\gamma}}{2m_{j}} \right]\\
	&=
	\frac{1}{2} \left[\sum_{i} q_{i} \hat{r}_{i,\alpha} \hat{r}_{i,\beta}, \sum_{j}\frac{\hat{p}_{j,\gamma} \hat{p}_{j,\gamma}}{2m_{j}} \right]\\
	&=
	\frac{1}{2} \sum_{i}\sum_{j} \left[ q_{i} \hat{r}_{i,\alpha} \hat{r}_{i,\beta}, \frac{\hat{p}_{j,\gamma} \hat{p}_{j,\gamma}}{2m_{j}} \right]\\
	&=
	\frac{1}{4}
	\sum_{i} \sum_{j} \frac{q_{i}}{m_{j}}
	\left[
	\hat{r}_{i,\alpha} \hat{r}_{i,\beta}
	,
	\hat{p}_{j,\gamma} \hat{p}_{j,\gamma}
	\right] \label{eq:Q_symmetric_last}
\end{align}
Now, in eq. \eqref{eq:Q_symmetric_last}, one can employ the following identities for a generic commutator $[AB, CD]$:
\begin{align}
	[AB, CD]
	&= A[B, C]D + [A, C]BD + CA[B, D] + C[A, D]B \label{eq:commutator_1}\\
	&= A[B, C]D + AC[B, D] + [A, C]DB + C[A, D]B \label{eq:commutator_2}
\end{align}
By using the first commutator itentity, eq. \eqref{eq:commutator_1}, one obtains:
\begin{align}
	&\frac{1}{2} \left[\sum_{i} q_{i} \hat{r}_{i,\alpha} \hat{r}_{i,\beta}, \hat{H}_{\text{mol}}\right]\\
	&=
	\frac{1}{4}
	\sum_{i} \sum_{j} \frac{q_{i}}{m_{j}}
	\left[
	\hat{r}_{i,\alpha} \hat{r}_{i,\beta}
	,
	\hat{p}_{j,\gamma} \hat{p}_{j,\gamma}
	\right]
	\\
	&=
	\frac{1}{4} \sum_{i} \sum_{j} \frac{q_{i}}{m_{j}}
	\left(
	\hat{r}_{i,\alpha} [\hat{r}_{i,\beta}, \hat{p}_{j,\gamma}]\hat{p}_{j,\gamma}
	+
	[\hat{r}_{i,\alpha}, \hat{p}_{j,\gamma}] \hat{r}_{i,\beta}\hat{p}_{j,\gamma}
	+
	\hat{p}_{j,\gamma} \hat{r}_{i,\alpha} [\hat{r}_{i,\beta}, \hat{p}_{j,\gamma}]
	+
	\hat{p}_{j,\gamma} [\hat{r}_{i,\alpha}, \hat{p}_{j,\gamma}]  \hat{r}_{i,\beta}
	\right)\\
	&=
	\frac{1}{4} \sum_{i} \sum_{j} \frac{q_{i}}{m_{j}}
	\left(
	\hat{r}_{i,\alpha} \iu\hbar\delta_{\beta\gamma} \hat{p}_{j,\gamma}
	+
	\iu\hbar\delta_{\alpha\gamma} \hat{r}_{i,\beta} \hat{p}_{j,\gamma}
	+
	\hat{p}_{j,\gamma} \hat{r}_{i,\alpha} \iu\hbar\delta_{\beta\gamma}
	+
	\hat{p}_{j,\gamma} \iu\hbar\delta_{\alpha\gamma}  \hat{r}_{i,\beta}
	\right) \delta_{ij} \\
	&=
	\frac{\iu\hbar}{4} \sum_{i} \frac{q_{i}}{m_{i}}
	\left(
	\hat{r}_{i,\alpha} \hat{p}_{i,\beta}
	+
	\hat{r}_{i,\beta} \hat{p}_{i,\alpha}
	+
	\hat{p}_{i,\beta} \hat{r}_{i,\alpha}
	+
	\hat{p}_{i,\alpha}  \hat{r}_{i,\beta}
	\right)
	\label{eq:form_1}
	\\
	&=
	\frac{\iu\hbar}{4} \sum_{i} \frac{q_{i}}{m_{i}}
	\left(
	\hat{r}_{i,\alpha} \hat{p}_{i,\beta}
	+
	\iu \hbar \delta_{\alpha\beta} + \hat{p}_{i,\alpha} \hat{r}_{i,\beta} 
	-
	\iu \hbar \delta_{\alpha\beta} + \hat{r}_{i,\alpha} \hat{p}_{i,\beta}
	+
	\hat{p}_{i,\alpha}  \hat{r}_{i,\beta}
	\right)\\
	&=
	\frac{\iu\hbar}{2} \sum_{i} \frac{q_{i}}{m_{i}}
	\left(
	\hat{r}_{i,\alpha} \hat{p}_{i,\beta}
	+
	\hat{p}_{i,\alpha} \hat{r}_{i,\beta}
	\right)
	\label{eq:form_2}
\end{align}
Eq. \eqref{eq:form_1}, presents an operator which is a $(\alpha,\beta)$-symmetric (Cartesian) tensor of rank 2 (as the traced electric-quadrupole moment). On the other hand, it is less obvious that the \virgolette{final} solution, eq. \eqref{eq:form_2} is a symmetric rank 2 tensor. As an alternative, by using the other \virgolette{expanded} form of the commutator $[AB, CD]$, i.e., eq. \eqref{eq:commutator_2}, one obtains:
\begin{align}
	&\frac{1}{2} \left[\sum_{i} q_{i} \hat{r}_{i,\alpha} \hat{r}_{i,\beta}, \hat{H}_{\text{mol}}\right]\\
	&=
	\frac{1}{4} \sum_{i} \sum_{j} \frac{q_{i}}{m_{j}}
	\left[ \hat{r}_{i,\alpha} \hat{r}_{i,\beta}, \hat{p}_{j,\gamma} \hat{p}_{j,\gamma} \right]\\
	&=
	\frac{1}{4} \sum_{i} \sum_{j} \frac{q_{i}}{m_{j}}
	\left(
	\hat{r}_{i,\alpha} [\hat{r}_{i,\beta}, \hat{p}_{j,\gamma}]\hat{p}_{j,\gamma}
	+
	\hat{r}_{i,\alpha} \hat{p}_{j,\gamma} [\hat{r}_{i,\beta}, \hat{p}_{j,\gamma}]
	+
	[\hat{r}_{i,\alpha}, \hat{p}_{j,\gamma}] \hat{p}_{j,\gamma} \hat{r}_{i,\beta}
	+
	\hat{p}_{j,\gamma} [\hat{r}_{i,\alpha}, \hat{p}_{j,\gamma}]  \hat{r}_{i,\beta}
	\right)\\
	&=
	\frac{1}{4} \sum_{i} \sum_{j} \frac{q_{i}}{m_{j}}
	\left(
	\hat{r}_{i,\alpha} \iu\hbar\delta_{\beta\gamma} \hat{p}_{j,\gamma}
	+
	\hat{r}_{i,\alpha}, \hat{p}_{j,\gamma} \iu\hbar\delta_{\beta\gamma}
	+
	\iu\hbar\delta_{\alpha\gamma} \hat{p}_{j,\gamma}, \hat{r}_{i,\beta}
	+
	\hat{p}_{j,\gamma} \iu\hbar\delta_{\alpha\gamma}  \hat{r}_{i,\beta}
	\right) \delta_{ij}
	\\
	&=
	\frac{\iu\hbar}{4}\sum_{i} \frac{q_{i}}{m_{i}}
	\left(
	\hat{r}_{i,\alpha} \hat{p}_{i,\beta}
	+
	\hat{r}_{i,\alpha} \hat{p}_{i,\beta}
	+
	\hat{p}_{i,\alpha} \hat{r}_{i,\beta}
	+
	\hat{p}_{i,\alpha} \hat{r}_{i,\beta}
	\right)\\
	&=
	\frac{\iu\hbar}{2}\sum_{i} \frac{q_{i}}{m_{i}}
	\left(
	\hat{r}_{i,\alpha} \hat{p}_{i,\beta}
	+
	\hat{p}_{i,\alpha} \hat{r}_{i,\beta}
	\right)
\end{align}
Which is identical to eq. \eqref{eq:form_2}. This solution is the one used by several authors \cite{Pedersen1995, Pedersen1999, Krykunov2006, Rizzo2006, Bernadotte2012, Vidal2015, Friese2016a} and it is the preferred one for this work. For the second term in the rhs of eq. \eqref{eq:Q_H_commutator} one obtains:
\begin{align}
	-\frac{1}{6} \left[ \sum_{i} q_{i} \hat{r}_{i,\gamma} \hat{r}_{i,\gamma} \delta_{\alpha\beta}, \hat{H}_{\text{mol}} \right]
	=
	-\frac{\iu\hbar}{6}
	\sum_{i} \frac{q_{i}}{m_{i}}
	\left( \hat{r}_{i,\gamma} \hat{p}_{i,\gamma} + \hat{p}_{i,\gamma} \hat{r}_{i,\gamma} \right) \delta_{\alpha\beta}
\end{align}
In conclusion:
\begin{align}
	[\oqu_{\alpha\beta}, \hat{H}_{\text{mol}}]
	&= \iu\oqu_{\alpha\beta}^{p},
\end{align}
where $\oqu_{\alpha\beta}^{p}$ is the traceless electric-quadrupole operator in velocity form, in atomic units ($\hbar = 1$):
\begin{equation}
	\oqu_{\alpha\beta}^{p}
	=
	\frac{1}{2}
	\sum_{i}
	\frac{q_{i}}{m_{i}}
	\left[
	\hat{r}_{i,\alpha} \hat{p}_{i,\beta}
	+
	\hat{p}_{i,\alpha} \hat{r}_{i,\beta}
	-\frac{1}{3}
	\left( \hat{r}_{i,\gamma} \hat{p}_{i,\gamma} + \hat{p}_{i,\gamma} \hat{r}_{i,\gamma} \right) \delta_{\alpha\beta}
	\right].
\end{equation}

\clearpage
\subsection{$[\omu_{\alpha}, \normalfont{\iu} \oqu_{\beta\gamma}^{p}]$} %
Here we derive the following commutator $[\omu_{\alpha}, \iu \oqu_{\beta\gamma}^{p}]$ for electronic wavefunctions:
\begin{align}
	[\omu_{\alpha}, \iu \oqu_{\beta\gamma}^{p}]
	&= \iu [\omu_{\alpha}, \oqu_{\beta\gamma}^{p}] \\
	&= \iu \left[\sum_{i} q_{i} \hat{r}_{i,\alpha},
	\frac{1}{2}
	\sum_{j}
	\frac{q_{j}}{m_{j}}
	\left(
	\hat{r}_{j,\beta} \hat{p}_{j,\gamma}
	+
	\hat{p}_{j,\beta} \hat{r}_{j,\gamma}
	-\frac{1}{3}
	\left( \hat{r}_{j,c} \hat{p}_{j,c} + \hat{p}_{j,c} \hat{r}_{j,c} \right) \delta_{\beta\gamma}
	\right)
	\right]\\
	&= \frac{\iu}{2} \left[\sum_{i} q_{i} \hat{r}_{i,\alpha},
	\sum_{j}
	\frac{q_{j}}{m_{j}}
	\left(
	\hat{r}_{j,\beta} \hat{p}_{j,\gamma}
	+
	\hat{p}_{j,\beta} \hat{r}_{j,\gamma}
	-\frac{1}{3}
	\left( \hat{r}_{j,c} \hat{p}_{j,c} + \hat{p}_{j,c} \hat{r}_{j,c} \right) \delta_{\beta\gamma}
	\right)
	\right]\\
	&= \frac{\iu}{2} \sum_{i} \sum_{j} \frac{q_{i}q_{j}}{m_{j}} 
	\left[ \hat{r}_{i,\alpha},
	\hat{r}_{j,\beta} \hat{p}_{j,\gamma}
	+
	\hat{p}_{j,\beta} \hat{r}_{j,\gamma}
	-\frac{1}{3}
	\left( \hat{r}_{j,c} \hat{p}_{j,c} + \hat{p}_{j,c} \hat{r}_{j,c} \right) \delta_{\beta\gamma}
	\right]\\
	&= \frac{\iu}{2} \sum_{i} \sum_{j} \frac{q_{i}q_{j}}{m_{j}} 
	\left(
	[\hat{r}_{i,\alpha},
	\hat{r}_{j,\beta} \hat{p}_{j,\gamma}]
	+
	[\hat{r}_{i,\alpha},
	\hat{p}_{j,\beta} \hat{r}_{j,\gamma}]
	-\frac{1}{3}
	[\hat{r}_{i,\alpha},
	\left( \hat{r}_{j,c} \hat{p}_{j,c} + \hat{p}_{j,c} \hat{r}_{j,c} \right) \delta_{\beta\gamma}]
	\right)\\
	&= \frac{\iu}{2} \sum_{i} \sum_{j} \frac{q_{i}q_{j}}{m_{j}} 
	\left(
	\hat{r}_{j,\beta} [\hat{r}_{i,\alpha},\hat{p}_{j,\gamma}]
	+
	[\hat{r}_{i,\alpha},
	\hat{p}_{j,\beta}]\hat{r}_{j,\gamma}
	-\frac{1}{3}
	\left(
	[\hat{r}_{i,\alpha},
	\hat{r}_{j,c} \hat{p}_{j,c}]\delta_{\beta\gamma}
	+
	[\hat{r}_{i,\alpha}, \hat{p}_{j,c} \hat{r}_{j,c}] \delta_{\beta\gamma}
	\right)
	\right)\\
	&= \frac{\iu}{2} \sum_{i} \sum_{j} \frac{q_{i}q_{j}}{m_{j}} 
	\left(
	\iu \hbar \hat{r}_{j,\beta}  \delta_{ij} \delta_{\alpha\gamma}
	+
	\iu \hbar \hat{r}_{j,\gamma} \delta_{ij} \delta_{\alpha\beta}
	-\frac{1}{3}
	\left(
	[\hat{r}_{i,\alpha},
	\hat{r}_{j,c} \hat{p}_{j,c}]\delta_{\beta\gamma}
	+
	[\hat{r}_{i,\alpha}, \hat{p}_{j,c} \hat{r}_{j,c}] \delta_{\beta\gamma}
	\right)
	\right)\\
	&= \frac{\iu}{2} \sum_{i} \sum_{j} \frac{q_{i}q_{j}}{m_{j}} 
	\left(
	\iu \hbar \hat{r}_{j,\beta}  \delta_{ij} \delta_{\alpha\gamma}
	+
	\iu \hbar \hat{r}_{j,\gamma} \delta_{ij} \delta_{\alpha\beta}
	-\frac{1}{3}
	\left(
	\iu \hbar \hat{r}_{j,\alpha} \delta_{ij} \delta_{\beta\gamma}
	+
	\iu \hbar \hat{r}_{j,\alpha} \delta_{ij} \delta_{\beta\gamma}
	\right)
	\right)\\
	&= \frac{\iu}{2} \sum_{i} \sum_{j} \frac{q_{i}q_{j}}{m_{j}} 
	\left(
	\iu \hbar \hat{r}_{j,\beta}  \delta_{\alpha\gamma}
	+
	\iu \hbar \hat{r}_{j,\gamma} \delta_{\alpha\beta}
	-\frac{\iu 2}{3} \hbar \hat{r}_{j,\alpha} \delta_{\beta\gamma}
	\right) \delta_{ij}\\
	&= \frac{\iu}{2} \sum_{i} \frac{q_{i}q_{i}}{m_{i}} 
	\left(
	\iu \hbar \hat{r}_{i,\beta}  \delta_{\alpha\gamma}
	+
	\iu \hbar \hat{r}_{i,\gamma} \delta_{\alpha\beta}
	-\frac{\iu 2}{3} \hbar \hat{r}_{i,\alpha} \delta_{\beta\gamma}
	\right) \\
	&= -\frac{1}{2} \sum_{i} \frac{q_{i}q_{i}}{m_{i}} 
	\left(
	\hat{r}_{i,\beta}  \delta_{\alpha\gamma}
	+
	\hat{r}_{i,\gamma} \delta_{\alpha\beta}
	-\frac{2}{3} \hat{r}_{i,\alpha} \delta_{\beta\gamma}
	\right) \\
	&=
	\frac{1}{2}
	\left(
	\vphantom{\frac{1}{1}}
	\omu_{\beta}\delta_{\alpha\gamma}
	+
	\omu_{\gamma}\delta_{\alpha\beta}
	\right)
	-
	\frac{1}{3} \omu_{\alpha}\delta_{\beta\gamma}
	\label{eq:comm_mu_iQ}
\end{align}
where, in the last step, we considered electronic wavefunctions and atomic units i.e., the index $i$ runs over the electrons only, and $q_{i} = -e = -1$ and $m_{i} = m_{e} = 1$.

\clearpage
\subsection{$[\oqu_{\alpha\beta}, \normalfont{\iu} \omu_{\gamma}^{p}]$} %
Here we derive the following commutator $[\oqu_{\alpha\beta}, \normalfont{\iu} \omu_{\gamma}^{p}]$ for electronic wavefunctions:
\begin{align}
	[\oqu_{\alpha\beta}, \iu\omu_{\gamma}^{p}]
	&=
	\iu [\oqu_{\alpha\beta}, \omu_{\gamma}^{p}]
	\\
	&=
	\iu
	\left[
	\frac{1}{2}\sum_{i}q_{i}
	\left(
	\hat{r}_{i,\alpha}\hat{r}_{i,\beta} -\frac{1}{3} \hat{r}_{i,c}\hat{r}_{i,c} \delta_{\alpha\beta}
	\right),
	\sum_{j} \frac{q_{j}}{m_{j}} \hat{p}_{j,\gamma}
	\right]
	\\
	&=
	\frac{\iu}{2}\sum_{i}\sum_{j} \frac{q_{i}q_{j}}{m_{j}}
	\left[
	\hat{r}_{i,\alpha}\hat{r}_{i,\beta} -\frac{1}{3} \hat{r}_{i,c}\hat{r}_{i,c} \delta_{\alpha\beta},
	\hat{p}_{j,\gamma}
	\right]
	\\
	&=
	\frac{\iu}{2}\sum_{i}\sum_{j} \frac{q_{i}q_{j}}{m_{j}}
	\left(
	\left[
	\hat{r}_{i,\alpha}\hat{r}_{i,\beta},
	\hat{p}_{j,\gamma}
	\right]
	- \frac{1}{3}
	\left[
	\hat{r}_{i,c}\hat{r}_{i,c},
	\hat{p}_{j,\gamma}
	\right]
	\delta_{\alpha\beta}
	\right)
	\\
	&=
	\frac{\iu}{2}\sum_{i}\sum_{j} \frac{q_{i}q_{j}}{m_{j}}
	\left(
	-\left[
	\hat{p}_{j,\gamma},
	\hat{r}_{i,\alpha}\hat{r}_{i,\beta}
	\right]
	+ \frac{1}{3}
	\left[
	\hat{p}_{j,\gamma},
	\hat{r}_{i,c}\hat{r}_{i,c}
	\right]
	\delta_{\alpha\beta}
	\right)
	\\
	&=
	\frac{\iu}{2}\sum_{i}\sum_{j} \frac{q_{i}q_{j}}{m_{j}}
	\left(
	+
	\iu \hbar \delta_{ji} \delta_{\gamma\alpha} \hat{r}_{i,\beta}
	+
	\iu \hbar \delta_{ji} \delta_{\gamma\beta} \hat{r}_{i,\alpha}
	- \frac{2}{3}
	\iu \hbar \delta_{ji} \delta_{\gamma c} \hat{r}_{i,c}
	\delta_{\alpha\beta}
	\right)
	\\
	&=
	-\frac{1}{2}\sum_{i}\sum_{j} \frac{q_{i}q_{j}}{m_{j}}
	\left(
	\delta_{\gamma\alpha} \hat{r}_{i,\beta}
	+
	\delta_{\gamma\beta} \hat{r}_{i,\alpha}
	- \frac{2}{3} \delta_{\gamma c} \hat{r}_{i,c}
	\delta_{\alpha\beta}
	\right)
	\delta_{ji}
	\\
	&=
	-\frac{1}{2}\sum_{i} \frac{q_{i}q_{i}}{m_{i}}
	\left(
	\delta_{\gamma\alpha} \hat{r}_{i,\beta}
	+
	\delta_{\gamma\beta} \hat{r}_{i,\alpha}
	- \frac{2}{3} \hat{r}_{i,\gamma}
	\delta_{\alpha\beta}
	\right)
	\\
	&=
	\frac{1}{2}
	\left(
	\vphantom{\frac{1}{1}}
	\omu_{\beta} \delta_{\gamma\alpha}
	+
	\omu_{\alpha} \delta_{\gamma\beta}
	\right)
	- \frac{1}{3}
	\omu_{\gamma} \delta_{\alpha\beta}
\end{align}
where, in the last step, we used the fact that we consider electronic wavefunctions and atomic units.

\clearpage
\section{Derivation of the velocity formulation} %
\label{sec:len_vel_derivation}

\subsection{$\bbeta$} %
By successive applications of eq. \eqref{eq:hyp_2}, one can re-write the pure electric-dipole first hyperpolarizability $\bbeta$ from the length, eq. \eqref{eq:beta}, to the velocity formulation:
\begin{align}
	\bbeta
	&= \pll \omu_{\alpha}; \omu_{\beta}, \omu_{\gamma} \prr\\
	&=
	\frac{1}{2\omega}
	\left(
	\pll [\omu_{\alpha}, \hat{H}_{\text{mol}}]; \omu_{\beta}, \omu_{\gamma} \prr_{\omega,\omega}
	+
	\pll [\omu_{\alpha},\omu_{\beta}]; \omu_{\gamma} \prr_{\omega}
	+
	\pll [\omu_{\alpha},\omu_{\gamma}]; \omu_{\beta} \prr_{\omega}
	\right)
	\\
	&=
	\frac{1}{2\omega}
	\pll \iu\omu_{\alpha}^{p}; \omu_{\beta}, \omu_{\gamma} \prr_{\omega,\omega}\\
	&=
	\frac{1}{2\omega}
	\pll \omu_{\beta}; \iu\omu_{\alpha}^{p}, \omu_{\gamma} \prr_{-2\omega,\omega}\\
	&=
	\frac{1}{2\omega}
	\left[
	-\frac{1}{\omega}
	\left(
	\pll [\omu_{\beta}, \hat{H}_{\text{mol}}]; \iu\omu_{\alpha}^{p}, \omu_{\gamma} \prr_{-2\omega,\omega}
	+
	\pll [\omu_{\beta}, \iu\omu_{\alpha}^{p}]; \omu_{\gamma} \prr_{\omega}
	+
	\pll [\omu_{\beta}, \omu_{\gamma}]; \iu\omu_{\alpha}^{p} \prr_{-2\omega}
	\right)
	\right]
	\\
	&=
	-\frac{1}{2\omega^{2}}
	\left(
	\pll \iu\omu_{\beta}^{p}; \iu\omu_{\alpha}^{p}, \omu_{\gamma} \prr_{-2\omega,\omega}
	+
	\pll -N_{e}\delta_{\beta\alpha}; \omu_{\gamma} \prr_{\omega}
	\right)\\
	&=
	-\frac{1}{2\omega^{2}}
	\pll \iu\omu_{\beta}^{p}; \iu\omu_{\alpha}^{p}, \omu_{\gamma} \prr_{-2\omega,\omega}\\
	&=
	-\frac{1}{2\omega^{2}}
	\pll \omu_{\gamma}; \iu\omu_{\alpha}^{p}, \iu\omu_{\beta}^{p} \prr_{-2\omega,\omega}\\
	&=
	-\frac{1}{2\omega^{2}}
	\left[
	-\frac{1}{\omega}
	\left(
	\pll [\omu_{\gamma}, \hat{H}_{\text{mol}}]; \iu\omu_{\alpha}^{p}, \iu\omu_{\beta}^{p} \prr_{-2\omega,\omega}
	+
	\pll [\omu_{\gamma}, \iu\omu_{\alpha}^{p}]; \iu\omu_{\beta}^{p} \prr_{\omega}
	+
	\pll [\omu_{\gamma}, \iu\omu_{\beta}^{p}]; \iu\omu_{\alpha}^{p} \prr_{-2\omega}
	\right)
	\right]
	\\
	&=
	\frac{1}{2\omega^{3}}
	\left(
	\pll  \iu\omu_{\gamma}^{p}; \iu\omu_{\alpha}^{p}, \iu\omu_{\beta}^{p} \prr_{-2\omega,\omega}
	+
	\pll -N_{e}\delta_{\gamma\alpha}; \iu\omu_{\beta}^{p} \prr_{\omega}
	+
	\pll -N_{e}\delta_{\gamma\beta}; \iu\omu_{\alpha}^{p} \prr_{-2\omega}
	\right)\\
	&=
	\frac{1}{2\omega^{3}}
	\pll  \iu\omu_{\gamma}^{p}; \iu\omu_{\alpha}^{p}, \iu\omu_{\beta}^{p} \prr_{-2\omega,\omega}\\
	&=
	\frac{1}{2\omega^{3}}
	\pll  \iu\omu_{\alpha}^{p}; \iu\omu_{\gamma}^{p}, \iu\omu_{\beta}^{p} \prr_{\omega,\omega}\\
	&=
	\frac{1}{2\omega^{3}}
	\pll  \iu\omu_{\alpha}^{p}; \iu\omu_{\beta}^{p}, \iu\omu_{\gamma}^{p} \prr_{\omega,\omega}\\
	&=
	-\frac{\iu}{2\omega^{3}}
	\pll  \omu_{\alpha}^{p}; \omu_{\beta}^{p}, \omu_{\gamma}^{p} \prr_{\omega,\omega}.
	\label{eq:beta_ppp}
\end{align}
\noindent
In the position representation [eqs. \eqref{eq:r_pos}-\eqref{eq:p_pos}], the electric-dipole moment in the velocity form, $\omu_{\alpha}^{p}$, is origin-independent [see eq. \eqref{eq:mu_p_gauge}], then $\beta_{\alpha\beta\gamma}$ is origin-independent also in the velocity formulation.

\clearpage
\subsection{$\aJ$} %
By proceeding in a similar fashion, for $\aJ$ one obtains:
\begin{align}
	\aJ
	&=
	\pll \oma_{\alpha}; \omu_{\beta}, \omu_{\gamma}\prr_{\omega,\omega} \\
	&=
	\pll \omu_{\beta}; \oma_{\alpha}, \omu_{\gamma}\prr_{-2\omega,\omega} \\
	&=
	-\frac{1}{\omega}
	\left(
	\pll [\omu_{\beta}, \hat{H}_{\text{mol}}]; \oma_{\alpha}, \omu_{\gamma} \prr_{-2\omega,\omega}
	+
	\pll [\omu_{\beta},\oma_{\alpha}]; \omu_{\gamma} \prr_{\omega}
	+
	\pll [\omu_{\beta},\omu_{\gamma}]; \oma_{\alpha} \prr_{-2\omega}\right)\\
	&=
	-\frac{1}{\omega}
	\left(
	\pll \iu\omu_{\beta}^{p}; \oma_{\alpha}, \omu_{\gamma} \prr_{-2\omega,\omega}
	+
	\pll [\omu_{\beta},\oma_{\alpha}]; \omu_{\gamma} \prr_{\omega}\right)\\
	&=
	-\frac{1}{\omega}
	\left(
	\pll \iu\omu_{\beta}^{p}; \oma_{\alpha}, \omu_{\gamma} \prr_{-2\omega,\omega}
	+
	\pll -\frac{\iu}{2c_{0}}\epsilon_{\beta\alpha c}\omu_{c}; \omu_{\gamma} \prr_{\omega} \right) \\
	&=
	-\frac{1}{\omega}
	\left(
	\pll \omu_{\gamma}; \oma_{\alpha}, \iu\omu_{\beta}^{p}  \prr_{-2\omega,\omega}
	+
	\pll -\frac{\iu}{2c_{0}}\epsilon_{\beta\alpha c}\omu_{c}; \omu_{\gamma} \prr_{\omega} \right) \\
	&=
	\left(-\frac{1}{\omega}\right)
	\left[
	-\frac{1}{\omega}
	\left(
	\pll [\omu_{\gamma}, \hat{H}_{\text{mol}}]; \oma_{\alpha}, \iu\omu_{\beta}^{p}  \prr_{-2\omega,\omega}
	+
	\pll [\omu_{\gamma}, \oma_{\alpha}]; , \iu\omu_{\beta}^{p}  \prr_{\omega}
	+
	\pll [\omu_{\gamma},  \iu\omu_{\beta}^{p}]; \oma_{\alpha}  \prr_{-2\omega}
	\right)
	+
	\pll -\frac{\iu}{2c_{0}}\epsilon_{\beta\alpha c}\omu_{c}; \omu_{\gamma} \prr_{\omega} \right] \\
	&=
	\left(-\frac{1}{\omega}\right)
	\left[
	-\frac{1}{\omega}
	\left(
	\pll \iu\omu_{\gamma}^{p}; \oma_{\alpha}, \iu\omu_{\beta}^{p}  \prr_{-2\omega,\omega}
	+
	\pll -\frac{\iu}{2c_{0}}\epsilon_{\gamma\alpha d}\omu_{d}; \iu\omu_{\beta}^{p}  \prr_{\omega}
	\right)
	+
	\pll -\frac{\iu}{2c_{0}}\epsilon_{\beta\alpha c}\omu_{c}; \omu_{\gamma} \prr_{\omega} \right] \\
	&=
	\left(-\frac{1}{\omega}\right)
	\left[
	-\frac{1}{\omega}
	\pll \iu\omu_{\gamma}^{p}; \oma_{\alpha}, \iu\omu_{\beta}^{p}  \prr_{-2\omega,\omega}
	+
	\frac{\iu}{2c_{0}\omega} \epsilon_{\gamma\alpha d} \pll \iu\omu_{\beta}^{p}; \omu_{d} \prr_{-\omega}
	-
	\frac{\iu}{2c_{0}} \epsilon_{\beta\alpha c} \pll \omu_{c}; \omu_{\gamma}\prr_{\omega}
	\right]
	\\
	&=
	\left(-\frac{1}{\omega}\right)
	\left[
	-\frac{1}{\omega}
	\pll \iu\omu_{\gamma}^{p}; \oma_{\alpha}, \iu\omu_{\beta}^{p}  \prr_{-2\omega,\omega}
	-
	\frac{\iu}{2c_{0}\omega} \epsilon_{\gamma\alpha d} \pll \iu\omu_{\beta}^{p}; \omu_{d} \prr_{\omega}
	-
	\frac{\iu}{2c_{0}} \epsilon_{\beta\alpha c} \pll \omu_{c}; \omu_{\gamma}\prr_{\omega}
	\right]
	\\
	&=
	\left(-\frac{1}{\omega}\right)
	\left[
	-\frac{1}{\omega}
	\pll \iu\omu_{\gamma}^{p}; \oma_{\alpha}, \iu\omu_{\beta}^{p}  \prr_{-2\omega,\omega}
	-
	\frac{\iu}{2c_{0}} \epsilon_{\gamma\alpha c} \pll \omu_{\beta}; \omu_{c} \prr_{\omega}
	-
	\frac{\iu}{2c_{0}} \epsilon_{\beta\alpha c} \pll \omu_{c}; \omu_{\gamma}\prr_{\omega}
	\right]\\
	&=
	\left(-\frac{1}{\omega}\right)
	\left[
	-\frac{1}{\omega}
	\pll \iu\omu_{\gamma}^{p}; \oma_{\alpha}, \iu\omu_{\beta}^{p}  \prr_{-2\omega,\omega}
	-
	\frac{\iu}{2c_{0}} \epsilon_{\gamma\alpha c} \pll \omu_{\beta}; \omu_{c} \prr_{\omega}
	-
	\frac{\iu}{2c_{0}} \epsilon_{\beta\alpha c} \pll \omu_{\gamma}; \omu_{c} \prr_{\omega}
	\right]\\
	&=
	\left(-\frac{1}{\omega}\right)
	\left[
	-\frac{1}{\omega}
	\pll \oma_{\alpha}; \iu\omu_{\gamma}^{p}, \iu\omu_{\beta}^{p}  \prr_{\omega,\omega}
	-
	\frac{\iu}{2c_{0}} \epsilon_{\gamma\alpha c} \pll \omu_{\beta}; \omu_{c} \prr_{\omega}
	-
	\frac{\iu}{2c_{0}} \epsilon_{\beta\alpha c} \pll \omu_{\gamma}; \omu_{c} \prr_{\omega}
	\right]\\
	&=
	\left(-\frac{1}{\omega}\right)
	\left[
	-\frac{1}{\omega}
	\pll \oma_{\alpha}; \iu\omu_{\beta}^{p}, \iu\omu_{\gamma}^{p}  \prr_{\omega,\omega}
	-
	\frac{\iu}{2c_{0}} \epsilon_{\gamma\alpha c} \pll \omu_{\beta}; \omu_{c} \prr_{\omega}
	-
	\frac{\iu}{2c_{0}} \epsilon_{\beta\alpha c} \pll \omu_{\gamma}; \omu_{c} \prr_{\omega}
	\right]\\
	&=
	\frac{1}{\omega^{2}}
	\pll \oma_{\alpha}; \iu\omu_{\beta}^{p}, \iu\omu_{\gamma}^{p}  \prr_{\omega,\omega}
	+\frac{\iu}{2c_{0}\omega}
	\left(\vphantom{\frac{1}{1}}
	\epsilon_{\gamma\alpha c} \pll \omu_{\beta}; \omu_{c} \prr_{\omega}
	+
	\epsilon_{\beta\alpha c} \pll \omu_{\gamma}; \omu_{c} \prr_{\omega}
	\right)
	\\
	&=
	-\frac{1}{\omega^{2}}
	\pll \oma_{\alpha}; \omu_{\beta}^{p}, \omu_{\gamma}^{p}  \prr_{\omega,\omega}
	+\frac{\iu}{2c_{0}\omega}
	\left(\vphantom{\frac{1}{1}}
	\epsilon_{\gamma\alpha c} \pll \omu_{\beta}; \omu_{c} \prr_{\omega}
	+
	\epsilon_{\beta\alpha c} \pll \omu_{\gamma}; \omu_{c} \prr_{\omega}
	\right).
	\label{eq:aJ_vel}
\end{align}
One can notice that the velocity formulation requires the computation of two tensors i.e., $\pll \oma_{\alpha}; \omu_{\beta}^{p}, \omu_{\gamma}^{p}  \prr_{\omega,\omega}$ and $\pll \omu_{\alpha}; \omu_{\beta} \prr_{\omega}$, the latter being the linear polarizability in length representation. Moreover, because the linear polarizability is origin-independent for any system (polar, neutral, or charged), the only origin-dependent tensor in this formulation is the first.

\clearpage
\subsection{$\bJ$} %
The mixed electric-dipole/magnetic-dipole first hyperpolarizability $\bJ$, eq. \eqref{eq:bJ}, contains two terms. For the first term in the rhs of eq. \eqref{eq:bJ} one obtains:
\begin{align}
	&\pll \omu_{\alpha}; \omu_{\beta}, \oma_{\gamma} \prr_{\omega,\omega}\\
	&=
	\frac{1}{2\omega}
	\left(
	\pll [\omu_{\alpha}, \hat{H}_{\text{mol}}]; \omu_{\beta}, \oma_{\gamma} \prr_{\omega,\omega}
	+
	\pll [\omu_{\alpha}, \omu_{\beta}]; \oma_{\gamma} \prr_{\omega}
	+
	\pll [\omu_{\alpha}, \oma_{\gamma}]; \omu_{\beta} \prr_{\omega}
	\right)\\
	&=
	\frac{1}{2\omega}
	\left(
	\pll \iu\omu_{\alpha}^{p}; \omu_{\beta}, \oma_{\gamma} \prr_{\omega,\omega}
	-\frac{\iu}{2c_{0}}\epsilon_{\alpha\gamma c}
	\pll \omu_{c}; \omu_{\beta} \prr_{\omega}
	\right)\\
	&=
	\frac{1}{2\omega}
	\left(
	\pll \omu_{\beta}; \iu\omu_{\alpha}^{p}, \oma_{\gamma} \prr_{-2\omega,\omega}
	-\frac{\iu}{2c_{0}}\epsilon_{\alpha\gamma c}
	\pll \omu_{c}; \omu_{\beta} \prr_{\omega}
	\right)\\
	&=
	\frac{1}{2\omega}
	\left[
	-\frac{1}{\omega}
	\left(
	\pll [\omu_{\beta}, \hat{H}_{\text{mol}}]; \iu\omu_{\alpha}^{p}, \oma_{\gamma} \prr_{-2\omega,\omega}
	+
	\pll [\omu_{\beta}, \iu\omu_{\alpha}^{p}]; \oma_{\gamma} \prr_{\omega}
	+
	\pll [\omu_{\beta}, \oma_{\gamma}]; \iu\omu_{\alpha}^{p} \prr_{-2\omega}
	\right)
	-\frac{\iu}{2c_{0}}\epsilon_{\alpha\gamma c}
	\pll \omu_{c}; \omu_{\beta} \prr_{\omega}
	\right]\\
	&=
	\frac{1}{2\omega}
	\left[
	-\frac{1}{\omega}
	\left(
	\pll \iu\omu_{\beta}^{p}; \iu\omu_{\alpha}^{p}, \oma_{\gamma} \prr_{-2\omega,\omega}
	-\frac{\iu}{2c_{0}}\epsilon_{\beta\gamma c}
	\pll \omu_{c}; \iu\omu_{\alpha}^{p} \prr_{-2\omega}
	\right)
	-\frac{\iu}{2c_{0}}\epsilon_{\alpha\gamma c}
	\pll \omu_{c}; \omu_{\beta} \prr_{\omega}
	\right]\\
	&=
	\frac{1}{2\omega}
	\left[
	-\frac{1}{\omega}
	\pll \iu\omu_{\beta}^{p}; \iu\omu_{\alpha}^{p}, \oma_{\gamma} \prr_{-2\omega,\omega}
	+\frac{\iu}{2c_{0}\omega}\epsilon_{\beta\gamma c}
	\pll \omu_{c}; \iu\omu_{\alpha}^{p} \prr_{-2\omega}
	-\frac{\iu}{2c_{0}}\epsilon_{\alpha\gamma c}
	\pll \omu_{c}; \omu_{\beta} \prr_{\omega}
	\right]\\
	&=
	\frac{1}{2\omega}
	\left[
	-\frac{1}{\omega}
	\pll \iu\omu_{\alpha}^{p}; \iu\omu_{\beta}^{p}, \oma_{\gamma} \prr_{\omega,\omega}
	+\frac{\iu}{2c_{0}\omega}\epsilon_{\beta\gamma c}
	\pll \iu\omu_{\alpha}^{p}; \omu_{c} \prr_{2\omega}
	-\frac{\iu}{2c_{0}}\epsilon_{\alpha\gamma c}
	\pll \omu_{c}; \omu_{\beta} \prr_{\omega}
	\right]\\
	&=
	\frac{1}{2\omega}
	\left[
	-\frac{1}{\omega}
	\pll \iu\omu_{\alpha}^{p}; \iu\omu_{\beta}^{p}, \oma_{\gamma} \prr_{\omega,\omega}
	+\frac{\iu}{c_{0}}\epsilon_{\beta\gamma c}
	\pll \omu_{\alpha}; \omu_{c} \prr_{2\omega}
	-\frac{\iu}{2c_{0}}\epsilon_{\alpha\gamma c}
	\pll \omu_{c}; \omu_{\beta} \prr_{\omega}
	\right]\\
	&=
	-\frac{1}{2\omega^{2}}
	\pll \iu\omu_{\alpha}^{p}; \iu\omu_{\beta}^{p}, \oma_{\gamma} \prr_{\omega,\omega}
	+\frac{\iu}{2c_{0}\omega}\epsilon_{\beta\gamma c}
	\pll \omu_{\alpha}; \omu_{c} \prr_{2\omega}
	-\frac{\iu}{4c_{0}\omega}\epsilon_{\alpha\gamma c}
	\pll \omu_{c}; \omu_{\beta} \prr_{\omega}\\
	&=
	-\frac{1}{2\omega^{2}}
	\pll \iu\omu_{\alpha}^{p}; \iu\omu_{\beta}^{p}, \oma_{\gamma} \prr_{\omega,\omega}
	+\frac{\iu}{2c_{0}\omega}
	\left(\vphantom{\frac{1}{1}}
	\epsilon_{\beta\gamma c}  \pll \omu_{\alpha}; \omu_{c} \prr_{2\omega}
	-
	\frac{1}{2}
	\epsilon_{\alpha\gamma c} \pll \omu_{c}; \omu_{\beta} \prr_{\omega}
	\right).
	\label{eq:bJ_vel}
\end{align}
And, as the second term in the rhs of eq. \eqref{eq:bJ} is obtained from the first by permuting the $(\beta,\gamma)$-indices, we can obtain its velocity formulation from the $(\beta \leftrightarrow \gamma)$ permutation of eq. \eqref{eq:bJ_vel}:
\begin{align}
	\pll \omu_{\alpha}; \omu_{\gamma}, \oma_{\beta} \prr_{\omega,\omega}
	=
	-\frac{1}{2\omega^{2}}
	\pll \iu\omu_{\alpha}^{p}; \iu\omu_{\gamma}^{p}, \oma_{\beta} \prr_{\omega,\omega}
	+\frac{\iu}{2c_{0}\omega}
	\left(\vphantom{\frac{1}{1}}
	\epsilon_{\gamma\beta c}  \pll \omu_{\alpha}; \omu_{c} \prr_{2\omega}
	-
	\frac{1}{2}
	\epsilon_{\alpha\beta c} \pll \omu_{c}; \omu_{\gamma} \prr_{\omega}
	\right),
\end{align}
and, by using the equality $\epsilon_{\gamma\beta c} = -\epsilon_{\beta\gamma c}$, the total sum reads:
\begin{align}
	\bJ 
	&=
	\frac{1}{2\omega^{2}}
	\left(\vphantom{\frac{1}{1}}
	\pll \omu_{\alpha}^{p}; \omu_{\beta}^{p}, \oma_{\gamma} \prr_{\omega,\omega}
	+
	\pll \omu_{\alpha}^{p}; \omu_{\gamma}^{p}, \oma_{\beta} \prr_{\omega,\omega}
	\right)
	-\frac{\iu}{4c_{0}\omega}
	\left(\vphantom{\frac{1}{1}}
	\epsilon_{\alpha\gamma c} \pll \omu_{c}; \omu_{\beta} \prr_{\omega}
	+
	\epsilon_{\alpha\beta c} \pll \omu_{c}; \omu_{\gamma} \prr_{\omega}
	\right).
\end{align}
As for $\aJ$, the computation of $\bJ$ in the velocity formulation requires the calculation of two tensors: $\pll \omu_{\alpha}^{p}; \omu_{\beta}^{p}, \oma_{\gamma} \prr_{\omega,\omega}$ and $\pll \omu_{\alpha}; \omu_{\beta} \prr_{\omega}$. Again, the origin-dependent tensor is only the first one, the second, being the linear polarizability, is intrinsically origin-independent.

\clearpage
\subsection{$\aK$} %
By proceeding in a similar fashion, for $\aK$ one obtains:
\begin{align}
	{}^{\alpha}K_{\alpha\beta\gamma\delta}
	&=
	\pll \oqu_{\alpha\delta}; \omu_{\beta}, \omu_{\gamma} \prr_{\omega,\omega}
	\\
	&=
	\frac{1}{2\omega}
	\left(
	\pll [\oqu_{\alpha\delta}, \hat{H}_{\text{mol}}]; \omu_{\beta}, \omu_{\gamma} \prr_{\omega,\omega}
	+
	\pll [\oqu_{\alpha\delta}, \omu_{\beta}], \omu_{\gamma} \prr_{\omega}
	+
	\pll [\oqu_{\alpha\delta}, \omu_{\gamma}], \omu_{\beta} \prr_{\omega}
	\right)
	\\
	&=
	\frac{1}{2\omega}
	\pll \iu\oqu_{\alpha\delta}^{p}; \omu_{\beta}, \omu_{\gamma} \prr_{\omega,\omega}
	\\
	&=
	\frac{1}{2\omega}
	\pll \omu_{\beta}; \iu\oqu_{\alpha\delta}^{p}, \omu_{\gamma} \prr_{-2\omega,\omega}
	\\
	&=
	\frac{1}{2\omega}
	\left[
	-\frac{1}{\omega}
	\left(
	\pll [\omu_{\beta}, \hat{H}_{\text{mol}}]; \iu\oqu_{\alpha\delta}^{p}, \omu_{\gamma} \prr_{-2\omega,\omega}
	+
	\pll [\omu_{\beta}, \iu\oqu_{\alpha\delta}^{p}]; \omu_{\gamma} \prr_{\omega}
	+
	\pll [\omu_{\beta}, \omu_{\gamma}]; \iu\oqu_{\alpha\delta}^{p}\prr_{-2\omega}
	\right)
	\right]
	\\
	&=
	-\frac{1}{2\omega^{2}}
	\left(
	\pll \iu \omu_{\beta}^{p}; \iu\oqu_{\alpha\delta}^{p}, \omu_{\gamma} \prr_{-2\omega,\omega}
	+
	\pll [\omu_{\beta}, \iu\oqu_{\alpha\delta}^{p}]; \omu_{\gamma} \prr_{\omega}
	\right)
	\\
	&=
	-\frac{1}{2\omega^{2}}
	\left(
	\pll \omu_{\gamma}; \iu\oqu_{\alpha\delta}^{p}, \iu \omu_{\beta}^{p} \prr_{-2\omega,\omega}
	+
	\pll [\omu_{\beta}, \iu\oqu_{\alpha\delta}^{p}]; \omu_{\gamma} \prr_{\omega}
	\right)
	\\
	&=
	-\frac{1}{2\omega^{2}}
	\left[
	-\frac{1}{\omega}
	\left(
	\pll [\omu_{\gamma}, \hat{H}_{\text{mol}}]; \iu\oqu_{\alpha\delta}^{p}, \iu \omu_{\beta}^{p} \prr_{-2\omega,\omega}
	+
	\pll
	[\omu_{\gamma}, \iu \oqu_{\alpha\delta}^{p}];
	\iu \omu_{\beta}^{p}
	\prr_{\omega}
	+
	\pll
	[\omu_{\gamma}, \iu \omu_{\beta}^{p}];
	\iu \oqu_{\alpha\delta}^{p} 
	\prr_{-2\omega}
	\right)
	+
	\pll [\omu_{\beta}, \iu\oqu_{\alpha\delta}^{p}]; \omu_{\gamma} \prr_{\omega}
	\right]
	\\
	&=
	-\frac{1}{2\omega^{2}}
	\left[
	-\frac{1}{\omega}
	\left(
	\pll \iu \omu_{\gamma}^{p}; \iu\oqu_{\alpha\delta}^{p}, \iu \omu_{\beta}^{p} \prr_{-2\omega,\omega}
	+
	\pll
	[\omu_{\gamma}, \iu \oqu_{\alpha\delta}^{p}];
	\iu \omu_{\beta}^{p}
	\prr_{\omega}
	\right)
	+
	\pll [\omu_{\beta}, \iu\oqu_{\alpha\delta}^{p}]; \omu_{\gamma} \prr_{\omega}
	\right]
	\\
	&=
	-\frac{1}{2\omega^{2}}
	\left[
	-\frac{1}{\omega}
	\left(
	\pll \iu \omu_{\beta}^{p}; \iu\oqu_{\alpha\delta}^{p}, \iu \omu_{\gamma}^{p} \prr_{-2\omega,\omega}
	+
	\pll
	[\omu_{\gamma}, \iu \oqu_{\alpha\delta}^{p}];
	\iu \omu_{\beta}^{p}
	\prr_{\omega}
	\right)
	+
	\pll [\omu_{\beta}, \iu\oqu_{\alpha\delta}^{p}]; \omu_{\gamma} \prr_{\omega}
	\right]
	\\
	&=
	-\frac{1}{2\omega^{2}}
	\left[
	-\frac{1}{\omega}
	\left(
	\pll \iu\oqu_{\alpha\delta}^{p};  \iu \omu_{\beta}^{p}, \iu \omu_{\gamma}^{p} \prr_{\omega,\omega}
	+
	\pll
	[\omu_{\gamma}, \iu \oqu_{\alpha\delta}^{p}];
	\iu \omu_{\beta}^{p}
	\prr_{\omega}
	\right)
	+
	\pll [\omu_{\beta}, \iu\oqu_{\alpha\delta}^{p}]; \omu_{\gamma} \prr_{\omega}
	\right]
	\\
	&=
	-\frac{1}{2\omega^{2}}
	\left[
	-\frac{1}{\omega}
	\pll \iu\oqu_{\alpha\delta}^{p};  \iu \omu_{\beta}^{p}, \iu \omu_{\gamma}^{p} \prr_{\omega,\omega}
	-\frac{1}{\omega}
	\pll
	[\omu_{\gamma}, \iu \oqu_{\alpha\delta}^{p}];
	\iu \omu_{\beta}^{p}
	\prr_{\omega}
	+
	\pll [\omu_{\beta}, \iu\oqu_{\alpha\delta}^{p}]; \omu_{\gamma} \prr_{\omega}
	\right]
	\\
	&=
	-\frac{1}{2\omega^{2}}
	\left[
	\frac{\iu}{\omega}
	\pll \oqu_{\alpha\delta}^{p}; \omu_{\beta}^{p}, \omu_{\gamma}^{p} \prr_{\omega,\omega}
	-\frac{1}{\omega}
	\pll
	[\omu_{\gamma}, \iu \oqu_{\alpha\delta}^{p}]; \iu\omu_{\beta}^{p}
	\prr_{\omega}
	+
	\pll
	[\omu_{\beta}, \iu\oqu_{\alpha\delta}^{p}]; \omu_{\gamma}
	\prr_{\omega}
	\right]
\end{align}
Let consider the last two terms in the previous equation. By using the commutator $[\omu_{\alpha}, \iu \oqu_{\beta\gamma}^{p}]$ given in eq. \eqref{eq:comm_mu_iQ}, the first term reads:
\begin{align}
	-\frac{1}{\omega}
	\pll
	[\omu_{\gamma}, \iu \oqu_{\alpha\delta}^{p}]; \omu_{\beta}
	\prr_{\omega}
	&=
	-\frac{1}{2\omega}
	\left(
	\vphantom{\frac{1}{1}}
	\pll
	\omu_{\alpha}; \iu\omu_{\beta}^{p}
	\prr_{\omega}
	\delta_{\gamma\delta}
	+
	\pll
	\omu_{\delta}; \iu\omu_{\beta}^{p}
	\prr_{\omega}
	\delta_{\gamma\alpha}
	\right)
	+
	\frac{1}{3\omega}
	\pll
	\omu_{\gamma}; \iu\omu_{\beta}^{p}
	\prr_{\omega}
	\delta_{\alpha\delta}
	\\
	&=
	-\frac{1}{2\omega}
	\left(
	\vphantom{\frac{1}{1}}
	\pll
	\iu\omu_{\beta}^{p}; \omu_{\alpha}
	\prr_{-\omega}
	\delta_{\gamma\delta}
	+
	\pll
	\iu\omu_{\beta}^{p}; \omu_{\delta}
	\prr_{-\omega}
	\delta_{\gamma\alpha}
	\right)
	+
	\frac{1}{3\omega}
	\pll
	\iu\omu_{\beta}^{p}; \omu_{\gamma}
	\prr_{-\omega}
	\delta_{\alpha\delta}
	\\
	&=
	\frac{1}{2}
	\left(
	\vphantom{\frac{1}{1}}
	\pll
	\omu_{\alpha}; \omu_{\beta}
	\prr_{\omega}
	\delta_{\gamma\delta}
	+
	\pll
	\omu_{\delta}; \omu_{\beta}
	\prr_{\omega}
	\delta_{\gamma\alpha}
	\right)
	-
	\frac{1}{3}
	\pll
	\omu_{\gamma}; \omu_{\beta}
	\prr_{\omega}
	\delta_{\alpha\delta}
\end{align}
and the second term is:
\begin{equation}
	\pll
	[\omu_{\beta}, \iu \oqu_{\alpha\delta}^{p}]; \omu_{\gamma}
	\prr_{\omega}
	=
	\frac{1}{2}
	\left(
	\vphantom{\frac{1}{1}}
	\pll
	\omu_{\alpha}; \omu_{\gamma}
	\prr_{\omega}
	\delta_{\beta\delta}
	+
	\pll
	\omu_{\delta}; \omu_{\gamma}
	\prr_{\omega}
	\delta_{\beta\alpha}
	\right)
	-
	\frac{1}{3}
	\pll
	\omu_{\beta}; \omu_{\gamma}
	\prr_{\omega}
	\delta_{\alpha\delta}
\end{equation}
So, at the end:
\begin{align}
	\begin{split}
		{}^{\alpha}K_{\alpha\beta\gamma\delta}
		=
		&-\frac{\iu}{2\omega^{3}}
		\pll
		\oqu_{\alpha\delta}^{p}; \omu_{\beta}^{p}, \omu_{\gamma}^{p} 
		\prr_{\omega, \omega}\\
		&
		-\frac{1}{4\omega^{2}}
		\biggl(
		\vphantom{\frac{1}{1}}
		\pll
		\omu_{\alpha}; \omu_{\beta}
		\prr_{\omega}
		\delta_{\gamma\delta}
		+
		\pll
		\omu_{\delta}; \omu_{\beta}
		\prr_{\omega}
		\delta_{\gamma\alpha}
		+
		\pll
		\omu_{\alpha}; \omu_{\gamma}
		\prr_{\omega}
		\delta_{\beta\delta}
		+
		\pll
		\omu_{\delta}; \omu_{\gamma}
		\prr_{\omega}
		\delta_{\beta\alpha}
		\biggr)\\
		&
		+\frac{1}{3\omega^{2}}
		\pll
		\omu_{\gamma}; \omu_{\beta}
		\prr_{\omega}
		\delta_{\alpha\delta}
	\end{split}
\end{align}
Consequently, from a computational point of view, the computation of $\aK$ in the velocity formulation requires the computation of two response tensors: $\pll \oqu_{\alpha\delta}^{p}; \omu_{\beta}^{p}, \omu_{\gamma}^{p} \prr_{\omega, \omega}$ and $\pll \omu_{\alpha}; \omu_{\beta} \prr_{\omega}$; the latter being the linear polarizability (in the length form). Again, one can notice that the origin-dependence is \virgolette{isolated} in the first term, i.e., the term containing the electric-quadrupole moment operator in velocity form.

\clearpage
\subsection{$\bK$} %
The mixed electric-dipole/electric-quadrupole first hyperpolarizability $\bK$, eq. \eqref{eq:bK}, contains two terms. In particular, the second term is obtained from the first by permuting the $(\beta,\gamma)$-indices. For the first term in the rhs of eq. \eqref{eq:bK} one obtains:
\begin{align}
	&\pll \omu_{\alpha}; \omu_{\beta}, \oqu_{\gamma\delta} \prr_{\omega, \omega}\\
	&=
	\frac{1}{2\omega}
	\left(
	\pll [\omu_{\alpha}, \hat{H}_{\text{mol}}]; \omu_{\beta}, \oqu_{\gamma\delta} \prr_{\omega, \omega}
	+
	\pll [\omu_{\alpha}, \omu_{\beta}]; \oqu_{\gamma\delta} \prr_{\omega}
	+
	\pll [\omu_{\alpha}, \oqu_{\gamma\delta}]; \omu_{\beta} \prr_{\omega}
	\right)\\
	&=
	\frac{1}{2\omega}
	\left(
	\pll \iu\omu_{\alpha}^{p}; \omu_{\beta}, \oqu_{\gamma\delta} \prr_{\omega, \omega}
	\right)\\
	&=
	\frac{1}{2\omega}
	\left(
	\pll \omu_{\beta}; \iu\omu_{\alpha}^{p}, \oqu_{\gamma\delta} \prr_{-2\omega, \omega}
	\right)\\
	&=
	\frac{1}{2\omega}
	\left[
	-\frac{1}{\omega}
	\left(
	\pll [\omu_{\beta}, \hat{H}_{\text{mol}}]; \iu\omu_{\alpha}^{p}, \oqu_{\gamma\delta} \prr_{-2\omega, \omega}
	+
	\pll [\omu_{\beta}, \iu\omu_{\alpha}^{p}]; \oqu_{\gamma\delta} \prr_{\omega}
	+
	\pll [\omu_{\beta}, \oqu_{\gamma\delta}]; \iu\omu_{\alpha}^{p} \prr_{-2\omega}
	\right)
	\right]\\
	&=
	-\frac{1}{2\omega^{2}}
	\pll \iu\omu_{\beta}^{p}; \iu\omu_{\alpha}^{p}, \oqu_{\gamma\delta} \prr_{-2\omega, \omega}
	\\
	&=
	-\frac{1}{2\omega^{2}}
	\pll \oqu_{\gamma\delta}; \iu\omu_{\alpha}^{p}, \iu\omu_{\beta}^{p} \prr_{-2\omega, \omega}
	\\
	&=
	-\frac{1}{2\omega^{2}}
	\left[
	-\frac{1}{\omega}
	\left(
	\pll [\oqu_{\gamma\delta}, \hat{H}_{\text{mol}}]; \iu\omu_{\alpha}^{p}, \iu\omu_{\beta}^{p} \prr_{-2\omega, \omega}
	+
	\pll [\oqu_{\gamma\delta}, \iu\omu_{\alpha}^{p}]; \iu\omu_{\beta}^{p} \prr_{\omega}
	+
	\pll [\oqu_{\gamma\delta}, \iu\omu_{\beta}^{p}]; \iu\omu_{\alpha}^{p} \prr_{-2\omega}
	\right)
	\right]
	\\
	&=
	\frac{1}{2\omega^{3}}
	\left(
	\pll
	\iu\oqu_{\gamma\delta}^{p}; \iu\omu_{\alpha}^{p}, \iu\omu_{\beta}^{p}
	\prr_{-2\omega, \omega}
	+
	\pll [\oqu_{\gamma\delta}, \iu\omu_{\alpha}^{p}]; \iu\omu_{\beta}^{p} \prr_{\omega}
	+
	\pll [\oqu_{\gamma\delta}, \iu\omu_{\beta}^{p}]; \iu\omu_{\alpha}^{p} \prr_{-2\omega}
	\right)\\
	\begin{split}
		&=
		\frac{1}{2\omega^{3}}
		\biggl[
		\pll
		\iu\oqu_{\gamma\delta}^{p}; \iu\omu_{\alpha}^{p}, \iu\omu_{\beta}^{p}
		\prr_{-2\omega, \omega}
		+\frac{1}{2}
		\left(
		\pll \omu_{\delta}; \iu\omu_{\beta}^{p} \prr_{\omega} \delta_{\alpha\gamma}
		+
		\pll \omu_{\gamma}; \iu\omu_{\beta}^{p} \prr_{\omega} \delta_{\alpha\delta}
		\right)
		-\frac{1}{3}
		\pll \omu_{\alpha}; \iu\omu_{\beta}^{p} \prr_{\omega} \delta_{\gamma\delta}
		\\
		& \qquad\qquad
		+\frac{1}{2}
		\left(\vphantom{\frac{1}{1}}
		\pll \omu_{\delta}; \iu\omu_{\alpha}^{p} \prr_{-2\omega} \delta_{\beta\gamma}
		+
		\pll \omu_{\gamma}; \iu\omu_{\alpha}^{p} \prr_{-2\omega} \delta_{\beta\delta}
		\right)
		-\frac{1}{3}
		\pll \omu_{\beta}; \iu\omu_{\alpha}^{p} \prr_{-2\omega} \delta_{\gamma\delta}
		\biggr]
	\end{split}\\
	\begin{split}
		&=
		\frac{1}{2\omega^{3}}
		\biggl[
		\pll
		\iu\omu_{\alpha}^{p}; \iu\omu_{\beta}^{p}, \iu\oqu_{\gamma\delta}^{p}
		\prr_{\omega, \omega}
		+\frac{1}{2}
		\left(
		\pll \iu\omu_{\beta}^{p}; \omu_{\delta} \prr_{-\omega} \delta_{\alpha\gamma}
		+
		\pll \iu\omu_{\beta}^{p}; \omu_{\gamma} \prr_{-\omega} \delta_{\alpha\delta}
		\right)
		-\frac{1}{3}
		\pll \iu\omu_{\beta}^{p}; \omu_{\alpha} \prr_{-\omega} \delta_{\gamma\delta}
		\\
		& \qquad\qquad
		+\frac{1}{2}
		\biggl[
		\left(\vphantom{\frac{1}{1}}
		\pll \iu\omu_{\alpha}^{p}; \omu_{\delta} \prr_{2\omega} \delta_{\beta\gamma}
		+
		\pll \iu\omu_{\alpha}^{p}; \omu_{\gamma} \prr_{2\omega} \delta_{\beta\delta}
		\right)
		-\frac{1}{3}
		\pll \iu\omu_{\alpha}^{p}; \omu_{\beta} \prr_{2\omega} \delta_{\gamma\delta}
		\biggr]
	\end{split}\\
	\begin{split}
		&=
		\frac{1}{2\omega^{3}}
		\pll
		\iu\omu_{\alpha}^{p}; \iu\omu_{\beta}^{p}, \iu\oqu_{\gamma\delta}^{p}
		\prr_{\omega, \omega}
		+ \frac{1}{2\omega^{2}}
		\biggl[
		-\frac{1}{2}
		\left(\vphantom{\frac{1}{1}}
		\pll \omu_{\beta}; \omu_{\delta} \prr_{\omega} \delta_{\alpha\gamma}
		+
		\pll \omu_{\beta}; \omu_{\gamma} \prr_{\omega} \delta_{\alpha\delta}
		\right)
		+\frac{1}{3}
		\pll \omu_{\beta}; \omu_{\alpha} \prr_{\omega} \delta_{\gamma\delta}
		\biggr]
		\\
		& \qquad\qquad
		+\frac{1}{\omega^{2}}
		\biggl[
		\frac{1}{2}
		\left(\vphantom{\frac{1}{1}}
		\pll \omu_{\alpha}; \omu_{\delta} \prr_{2\omega} \delta_{\beta\gamma}
		+
		\pll \omu_{\alpha}; \omu_{\gamma} \prr_{2\omega} \delta_{\beta\delta}
		\right)
		-\frac{1}{3}
		\pll \omu_{\alpha}; \omu_{\beta} \prr_{2\omega} \delta_{\gamma\delta}
		\biggr]
	\end{split}
\end{align}
The second term in the rhs of eq. \eqref{eq:bK} is obtained from the first by the permutation of the $(\beta,\gamma)$-indices. Consequently, we can derive the velocity formulation for the second term can be obtained by the permutation of the $(\beta,\gamma)$-indices:
\begin{equation}
	\begin{split}
		\pll \omu_{\alpha}; \omu_{\gamma}, \oqu_{\beta\delta} \prr_{\omega, \omega}
		&=
		\frac{1}{2\omega^{3}}
		\pll
		\iu\omu_{\alpha}^{p}; \iu\omu_{\gamma}^{p}, \iu\oqu_{\beta\delta}^{p}
		\prr_{\omega, \omega}
		+ \frac{1}{2\omega^{2}}
		\biggl[
		-\frac{1}{2}
		\left(\vphantom{\frac{1}{1}}
		\pll \omu_{\gamma}; \omu_{\delta} \prr_{\omega} \delta_{\alpha\beta}
		+
		\pll \omu_{\gamma}; \omu_{\beta} \prr_{\omega} \delta_{\alpha\delta}
		\right)
		+\frac{1}{3}
		\pll \omu_{\gamma}; \omu_{\alpha} \prr_{\omega} \delta_{\beta\delta}
		\biggr]
		\\
		& \qquad\qquad
		+\frac{1}{\omega^{2}}
		\biggl[
		\frac{1}{2}
		\left(\vphantom{\frac{1}{1}}
		\pll \omu_{\alpha}; \omu_{\delta} \prr_{2\omega} \delta_{\gamma\beta}
		+
		\pll \omu_{\alpha}; \omu_{\beta} \prr_{2\omega} \delta_{\gamma\delta}
		\right)
		-\frac{1}{3}
		\pll \omu_{\alpha}; \omu_{\gamma} \prr_{2\omega} \delta_{\beta\delta}
		\biggr]
	\end{split}
\end{equation}
Finally, the mixed electric-dipole/electric-quadrupole first hyperpolarizability $\bK$ can be written in the velocity formulation as:
\begin{align}
	\begin{split}
		\bK &=
		-\frac{\iu}{2\omega^{3}}
		\left(
		\pll \omu_{\alpha}^{p}; \omu_{\beta}^{p}, \oqu_{\gamma\delta}^{p} \prr_{\omega, \omega}
		+
		\pll \omu_{\alpha}^{p}; \omu_{\gamma}^{p}, \oqu_{\beta\delta}^{p} \prr_{\omega, \omega}
		\right)\\
		&\qquad
		+\frac{1}{2\omega^{2}}
		\biggl[
		-\frac{1}{2}
		\left(\vphantom{\frac{1}{1}}
		\pll \omu_{\beta}; \omu_{\delta} \prr_{\omega} \delta_{\alpha\gamma}
		+
		\pll \omu_{\beta}; \omu_{\gamma} \prr_{\omega} \delta_{\alpha\delta}
		\right)
		+\frac{1}{3}
		\pll \omu_{\beta}; \omu_{\alpha} \prr_{\omega} \delta_{\gamma\delta}
		\biggr]\\
		&\qquad
		+\frac{1}{\omega^{2}}
		\biggl[
		\frac{1}{2}
		\left(\vphantom{\frac{1}{1}}
		\pll \omu_{\alpha}; \omu_{\delta} \prr_{2\omega} \delta_{\beta\gamma}
		+
		\pll \omu_{\alpha}; \omu_{\gamma} \prr_{2\omega} \delta_{\beta\delta}
		\right)
		-\frac{1}{3}
		\pll \omu_{\alpha}; \omu_{\beta} \prr_{2\omega} \delta_{\gamma\delta}
		\biggr]\\
		&\qquad
		+\frac{1}{2\omega^{2}}
		\biggl[
		-\frac{1}{2}
		\left(\vphantom{\frac{1}{1}}
		\pll \omu_{\gamma}; \omu_{\delta} \prr_{\omega} \delta_{\alpha\beta}
		+
		\pll \omu_{\gamma}; \omu_{\beta} \prr_{\omega} \delta_{\alpha\delta}
		\right)
		+\frac{1}{3}
		\pll \omu_{\gamma}; \omu_{\alpha} \prr_{\omega} \delta_{\beta\delta}
		\biggr]\\
		&\qquad
		+\frac{1}{\omega^{2}}
		\biggl[
		\frac{1}{2}
		\left(\vphantom{\frac{1}{1}}
		\pll \omu_{\alpha}; \omu_{\delta} \prr_{2\omega} \delta_{\gamma\beta}
		+
		\pll \omu_{\alpha}; \omu_{\beta} \prr_{2\omega} \delta_{\gamma\delta}
		\right)
		-\frac{1}{3}
		\pll \omu_{\alpha}; \omu_{\gamma} \prr_{2\omega} \delta_{\beta\delta}
		\biggr]
	\end{split}\\
	\begin{split}
	&=
	-\frac{\iu}{2\omega^{3}}
	\left(
	\pll \omu_{\alpha}^{p}; \omu_{\beta}^{p}, \oqu_{\gamma\delta}^{p} \prr_{\omega,\omega}
	+
	\pll \omu_{\alpha}^{p}; \omu_{\gamma}^{p}, \oqu_{\beta\delta}^{p} \prr_{\omega,\omega}
	\right)\\&\qquad
	+ \frac{1}{2\omega^{2}}
	\left[
	-\frac{1}{2}
	\left( \vphantom{\frac{1}{1}}
	\pll \omu_{\beta}; \omu_{\delta} \prr_{\omega} \delta_{\alpha\gamma}
	+
	\pll \omu_{\gamma}; \omu_{\delta} \prr_{\omega} \delta_{\alpha\beta}
	+
	2
	\pll \omu_{\beta}; \omu_{\gamma} \prr_{\omega} \delta_{\alpha\delta}
	\right)
	+ \frac{1}{3}
	\left(\vphantom{\frac{1}{1}}
	\pll \omu_{\beta}; \omu_{\alpha} \prr_{\omega} \delta_{\gamma\delta}
	+
	\pll \omu_{\gamma}; \omu_{\alpha} \prr_{\omega} \delta_{\beta\delta}
	\right)
	\right]\\&\qquad
	+\frac{1}{\omega^{2}}
	\left[
	\vphantom{\frac{1}{1}}
	\pll \omu_{\alpha}; \omu_{\delta}\prr_{2\omega} \delta_{\beta\gamma}
	+
	\frac{1}{6}
	\left(
	\vphantom{\frac{1}{1}}
	\pll \omu_{\alpha}; \omu_{\gamma} \prr_{2\omega} \delta_{\beta\delta}
	+
	\pll \omu_{\alpha}; \omu_{\beta} \prr_{2\omega} \delta_{\gamma\delta}
	\right)
	\right]
	\end{split}\\
	\begin{split}
		&=
		-\frac{\iu}{2\omega^{3}}
		\left(
		\pll \omu_{\alpha}^{p}; \omu_{\beta}^{p}, \oqu_{\gamma\delta}^{p} \prr_{\omega,\omega}
		+
		\pll \omu_{\alpha}^{p}; \omu_{\gamma}^{p}, \oqu_{\beta\delta}^{p} \prr_{\omega,\omega}
		\right)\\&\qquad
		+ \frac{1}{12\omega^{2}}
		\left[
		-3
		\left( \vphantom{\frac{1}{1}}
		\pll \omu_{\beta}; \omu_{\delta} \prr_{\omega} \delta_{\alpha\gamma}
		+
		\pll \omu_{\gamma}; \omu_{\delta} \prr_{\omega} \delta_{\alpha\beta}
		+
		2
		\pll \omu_{\beta}; \omu_{\gamma} \prr_{\omega} \delta_{\alpha\delta}
		\right)
		+ 2
		\left(\vphantom{\frac{1}{1}}
		\pll \omu_{\beta}; \omu_{\alpha} \prr_{\omega} \delta_{\gamma\delta}
		+
		\pll \omu_{\gamma}; \omu_{\alpha} \prr_{\omega} \delta_{\beta\delta}
		\right)
		\right]\\&\qquad
		+\frac{1}{6\omega^{2}}
		\left[
		6
		\pll \omu_{\alpha}; \omu_{\delta}\prr_{2\omega} \delta_{\beta\gamma}
		+
		\left(
		\vphantom{\frac{1}{1}}
		\pll \omu_{\alpha}; \omu_{\gamma} \prr_{2\omega} \delta_{\beta\delta}
		+
		\pll \omu_{\alpha}; \omu_{\beta} \prr_{2\omega} \delta_{\gamma\delta}
		\right)
		\right]
	\end{split}\\
\begin{split}
	&=
	-\frac{\iu}{2\omega^{3}}
	\left(
	\pll \omu_{\alpha}^{p}; \omu_{\beta}^{p}, \oqu_{\gamma\delta}^{p} \prr_{\omega,\omega}
	+
	\pll \omu_{\alpha}^{p}; \omu_{\gamma}^{p}, \oqu_{\beta\delta}^{p} \prr_{\omega,\omega}
	\right)\\&\qquad
	+ \frac{1}{12\omega^{2}}
	\biggl[
	-3
	\left( \vphantom{\frac{1}{1}}
	\pll \omu_{\beta}; \omu_{\delta} \prr_{\omega} \delta_{\alpha\gamma}
	+
	\pll \omu_{\gamma}; \omu_{\delta} \prr_{\omega} \delta_{\alpha\beta}
	+
	2
	\pll \omu_{\beta}; \omu_{\gamma} \prr_{\omega} \delta_{\alpha\delta}
	\right)
	+ 2
	\left(\vphantom{\frac{1}{1}}
	\pll \omu_{\beta}; \omu_{\alpha} \prr_{\omega} \delta_{\gamma\delta}
	+
	\pll \omu_{\gamma}; \omu_{\alpha} \prr_{\omega} \delta_{\beta\delta}
	\right)
	\\&\qquad
	+12
	\pll \omu_{\alpha}; \omu_{\delta}\prr_{2\omega} \delta_{\beta\gamma}
	+2
	\biggl(
	\vphantom{\frac{1}{1}}
	\pll \omu_{\alpha}; \omu_{\gamma} \prr_{2\omega} \delta_{\beta\delta}
	+
	\pll \omu_{\alpha}; \omu_{\beta} \prr_{2\omega} \delta_{\gamma\delta}
	\biggr)
	\biggr]
\end{split}
\end{align}
In this case, the velocity formulation of $\bK$ requires the computation of three tensors: $\pll \omu_{\alpha}^{p}; \omu_{\beta}^{p}, \oqu_{\gamma\delta}^{p} \prr_{\omega, \omega}$, and the linear electric-dipole polarizability computed at $\omega$ and $2\omega$: $\pll \omu_{\alpha}; \omu_{\beta} \prr_{\omega}$, and $\pll \omu_{\alpha}; \omu_{\beta} \prr_{2\omega}$. However, similarly to the previous cases, the origin-dependence of this formulation is isolated in $\pll \omu_{\alpha}^{p}; \omu_{\beta}^{p}, \oqu_{\gamma\delta}^{p} \prr_{\omega, \omega}$.

\clearpage
\subsection{$\alpha_{\alpha\beta}$} %
\label{sec:alpha_velocity}
By using the hypervirial relation for linear response functions, eq. \eqref{eq:hyp_1}, we derive the velocity formulation for the linear polarizability $\alpha_{\alpha\beta} = - \pll \omu_{\alpha}; \omu_{\beta} \prr_{\omega}$:
\begin{align}
	\pll \omu_{\alpha}; \omu_{\beta} \prr_{\omega}
	&=
	\frac{1}{\omega}
	\left(
	\pll [\omu_{\alpha}, \hat{H}_{\text{mol}}]; \omu_{\beta} \prr_{\omega}
	+
	\braket{0|[\omu_{\alpha}, \omu_{\beta}]|0}
	\right)
	\\
	&=
	\frac{1}{\omega}
	\left(
	\pll \iu \omu_{\alpha}^{p}; \omu_{\beta} \prr_{\omega}
	\right)
	\\
	&=
	\frac{1}{\omega}
	\left(
	\pll \omu_{\beta}; \iu \omu_{\alpha}^{p} \prr_{-\omega}
	\right)
	\\
	&=
	\frac{1}{\omega}
	\left[
	-\frac{1}{\omega}
	\left(
	\pll [\omu_{\beta}, \hat{H}_{\text{mol}}]; \iu \omu_{\alpha}^{p} \prr_{-\omega}
	+
	\braket{0|[\omu_{\beta}, \iu \omu_{\alpha}^{p}]|0}
	\right)
	\right]
	\\
	&=
	\frac{1}{\omega}
	\left[
	-\frac{1}{\omega}
	\left(
	\pll \iu \omu_{\beta}^{p}; \iu \omu_{\alpha}^{p} \prr_{-\omega}
	+
	\iu\braket{0|[\omu_{\beta}, \omu_{\alpha}^{p}]|0}
	\right)
	\right]
	\\
	&=
	\frac{1}{\omega}
	\left[
	-\frac{1}{\omega}
	\left(
	\pll \iu \omu_{\alpha}^{p}; \iu \omu_{\beta}^{p} \prr_{\omega}
	+
	\iu\braket{0|[\omu_{\beta}, \omu_{\alpha}^{p}]|0}
	\right)
	\right]
	\\
	&=
	\frac{1}{\omega}
	\left[
	-\frac{1}{\omega}
	\left(
	-\pll \omu_{\alpha}^{p}; \omu_{\beta}^{p} \prr_{\omega}
	+
	\iu (\iu N_{e} \delta_{\beta\alpha})
	\right)
	\right]
	\\
	&=
	\frac{1}{\omega}
	\left[
	-\frac{1}{\omega}
	\left(
	-\pll \omu_{\alpha}^{p}; \omu_{\beta}^{p} \prr_{\omega}
	-
	N_{e} \delta_{\beta\alpha}
	\right)
	\right]
	\\
	&=
	\frac{1}{\omega}
	\left[
	\frac{1}{\omega}
	\left(
	\pll \omu_{\alpha}^{p}; \omu_{\beta}^{p} \prr_{\omega}
	+
	N_{e} \delta_{\beta\alpha}
	\right)
	\right]
	\\
	&=
	\frac{1}{\omega^2}
	\left(
	\pll \omu_{\alpha}^{p}; \omu_{\beta}^{p} \prr_{\omega}
	+
	N_{e} \delta_{\beta\alpha}
	\right) \label{eq:ee_pp_Ne}
\end{align}
%
